\documentclass[12pt,a4paper]{article}

\usepackage{soul}        

\usepackage{amsmath,amssymb,epsf,epsfig}
\usepackage{array}
\usepackage{fancybox}
\allowdisplaybreaks[1]

\usepackage{tocbibind}
\usepackage{lipsum}
\usepackage[usenames]{color}
\usepackage{amsfonts,bm}
\usepackage{graphicx}
\usepackage{enumerate}
\usepackage{tikz}
\usetikzlibrary{arrows}
\usepackage{parskip}     
\usepackage[
      colorlinks=true,
      linkcolor=black,
      urlcolor=blue,
      filecolor=blue,
      citecolor=black,
      pdfstartview=FitH,
      pdftitle={},
      pdfauthor={},
      pdfsubject={},
      pdfkeywords={},
      pdfpagemode=UseNone,
      bookmarksopen=true,
      ]{hyperref}
\usepackage[all]{hypcap}     

\numberwithin{equation}{section}

\newcommand{\be}{\begin{equation}}
\newcommand{\ee}{\end{equation}}
\newcommand{\bea}{\begin{eqnarray}\displaystyle}
\newcommand{\eea}{\end{eqnarray}}

\def\beq{\begin{equation}}
\def\eeq{\end{equation}}
\def\beqa{\begin{eqnarray}}
\def\eeqa{\end{eqnarray}}
\def\bet{\begin{tabular}}
\def\eet{\end{tabular}}
\def\bs{\begin{split}}
\def\es{\end{split}}

\def\k{\kappa}

\def\one{{\hbox{\kern+.5mm 1\kern-.8mm l}}}
\def\zero{{\hbox{0\kern-1.5mm 0}}}

\definecolor{orange}{rgb}{1,0.5,0}

\newcommand{\bra}[1]{{\langle {#1} |\,}}
\newcommand{\ket}[1]{{\,| {#1} \rangle}}
\newcommand{\braket}[2]{\ensuremath{\langle #1 | #2 \rangle}}

\newcommand{\bean}{\begin{eqnarray*}}
\newcommand{\eean}{\end{eqnarray*}}

\begin{document}

\begin{center}
{\huge{ $tt^{*}$ Geometry of Modular Curves } }
\end{center}

\vspace{2cm}

\begin{center}
{\large{Riccardo Bergamin}}
\end{center}

\vspace{0.5 cm}

\begin{center}
{\large{SISSA, via Bonomea 265, I-34100 Trieste, ITALY}}
\end{center}

\vspace{2cm}

\begin{center}
{\large{
Abstract} }
\end{center}

Motivated by Vafa's model, we study the $tt^{*}$ geometry of a degenerate class of fractional quantum Hall effect (FQHE) models with an abelian group of symmetry acting transitively on the classical vacua. Despite it is not relevant for the phenomenology of the FQHE, this class of theories has interesting mathematical properties. We find that these models are parametrized by the family of modular curves $Y_{1}(N)= \mathbb{H}/\Gamma_{1}(N)$, labelled by an integer $N\geq 2$. Each point of the space of level $N$ is in correspondence with a one dimensional $\mathcal{N}=4$ Landau-Ginzburg theory, which is defined on an elliptic curve with $N$ vacua and $N$ poles in the fundamental cell. The modular curve $Y(N)= \mathbb{H}/\Gamma(N)$ is a cover of degree $N$ of $Y_{1}(N)$ and plays the role of spectral cover for the space of models. The presence of an abelian symmetry allows to diagonalize the Berry's connection of the vacuum bundle and the $tt^{*}$ equations turn out to be the well known $\hat{A}_{N-1}$ Toda equations. The underlying structure of the modular curves and the connection between geometry and number theory emerge clearly when we study the modular properties and classify the critical limits of these models.

\newpage

\tableofcontents

\section{Introduction}

Despite its discovery dates back more than thirty years ago \cite{discovery}, the physics of the fractional quantum Hall effect (FQHE) is not yet fully understood. One of the main open questions is the nature of the charged excitations of the Hall fluid. These quasi-holes are believed to possess anyonic statistics, but saying if they are abelian or non-abelian particles still represents a challenging problem for both theorists and experimentalists. From the theoretical point of view, many models have been developed to explain the observed filling fractions. Among these, the Laughlin's proposal \cite{ab1} and the idea of hierarchy states of Haldane and Halperin \cite{ab2,ab3}, as well as Jain's composite fermion theory \cite{ab4}, predict abelian anyonic statistics for the principal series of FQHE. Also the possibility of non-abelian statistics has been explored by several models for other filling fractions \cite{nab1,nab2,nab3,nab4}.\\ More recently, C.Vafa proposed in \cite{rif1} a unifying model of FQHE which leads to new predictions for the statistics of the quasi-holes. He shows that the effective theory of the FQH systems can be realized in a string theoretical context as $SL(2,\mathbb{C})$ Chern-Simons theory in the $2+1$ dimensional bulk. This construction motivates in addition a microscopic description in terms of a $\mathcal{N}=4$ supersymmetric Hamiltonian. More precisely, the prototype of one-particle supersymmetric model which is relevant for the FQHE physics is given by the Landau-Ginzburg theory with superpotential

\begin{equation}\label{class}   
W(z)= \sum_{\zeta \in L} e(\zeta)\log(z-\zeta),
\end{equation}

where $L$ is a discrete set of $\mathbb{C} $ and $e(\zeta)$ are real numbers. The variable $z$ is interpreted as the electron coordinate and the term $e(\zeta)\log(z-\zeta)$ is the two dimensional coulombic potential which describes the interaction between the electron and an external charge. In the standard setting of the FQH systems the source of electrostatic interaction is taken to be $L= \Lambda \cup S$, with $\Lambda$ a lattice and $S$ a set of positions of quasi-holes. The effect of the lattice is to reproduce the constant macroscopic magnetic field with $e(\lambda)=1$ units of magnetic flux at a point $\lambda \in \Lambda$. At this level the expression of the superpotential is just symbolic, since the sum is taken over an infinite set of points and the function is multi-valued. Therefore it requires a more precise definition according to the class of models that one is considering.\\ One of the main advantage of this model is that we do not need to compute the wave functions to describe the ground states, since these are labelled by operators in the chiral ring of the theory $\mathcal{R}= \frac{C[z]}{\partial W}$ \cite{rif4}. For a generic choice of the parameters defining $W(z)$, the classical vacua are isolated, i.e. the theory is massive, and the elements in the chiral ring are identified with their set of values at the critical points. Moreover, the Berry's connection of the vacuum bundle satisfies a set of equations known as $tt^{*}$ geometry. To have an answer about the statistics of the quasi-holes one needs to solve these equations and compute the Berry's connection for the ground states. 
However, finding the classical vacua and studying the $tt^{*}$ geometry of these models is rather complicated, unless one arranges the set of quasi-holes in some special configuration to have an enhancement of symmetry. In this way we can construct degenerate models of FQHE which are not realistic for phenomenological purposes, but at least analitically treatable.\\ In this paper we study a particular class of theories of this type which have an abelian subgroup of symmetry acting transitively on the set of vacua. This is the most convenient limit, since in this case the Berry's connection can be completely diagonalized in a basis of eigenstates of such symmetry and the $tt^{*}$ equations can be derived. It turns out that these models are parametrized modulo isomorphisms by the family of Riemann surfaces $Y_{1}(N)= \mathbb{H}/\Gamma_{1}(N)$, labelled by an integer $N\geq 2$, also known as modular curves for the congruence subgroup $\Gamma_{1}(N)$ of the modular group $SL(2,\mathbb{Z})$ \cite{rif2}. Each point of the curve of level $N$ identifies a theory where $\partial_{z}W(z)$ is an ellitpic function with a $\mathbb{Z}_{N}$ symmetry generated by a torsion point of the elliptic curve $\mathbb{C}/\Lambda$. By adding also the generators of the lattice one obtains a basis for the abelian symmetry group of the model. It is clear from such properties that, despite they are inspired by the FQHE setting, these models do not really describe that kind of physics. Indeed the charges of the quasi-holes inside the fundamental cell have to cancel the flux of the magnetic field in order to have a doubly periodic physics. What is really worth to study in this class of theories is the rich underlying mathematical structure.\\ For each connected component of the space of models one can define its spectral cover as the complex manifold whose points identify a model and a vacuum \cite{rif12}. These are the modular curves $Y(N)= \mathbb{H}/\Gamma(N)$ for the principal congruence subgroup $ \Gamma(N)$. More precisley, $Y(N)$ is a cover of $Y_{1}(N)$ of degree $N$, which is the number of vacua in the fundamental cell of the torus. The $tt^{*}$ equations simplify considerably on the spectral curve and can be normalized in the form of $\hat{A}_{N-1}$ Toda equations \cite{rif4}. These appear in all the models with a $\mathbb{Z}_{N}$ symmetry group which is transitive on the vacua. The modular curves are manifolds with cusps, which represent physically the RG flow fixed points of the theory. These are in correspondence with the equivalence classes of rationals with respect to the congruence subgroups. An outstandig fact is that for a given $N$ all the $\hat{A}_{Q-1}$ models with $Q \vert N$ are embedded in this class of theories as critical limits, providing the regularity conditions for the solutions to the equations. An exception is the case of $N=4$, where only $\hat{A}_{3}$ models appear. The beauty of the modular curves is that they possess various geometrical structures. For instance, in the modular curves $Y(N)$ of level $N=3,4,5 $ the cusps are located at the vertices of platonic solids inscribed in the Riemann sphere.\\ The theory of modular curves is also strictly related to number theory. These surfaces can be seen as projective algebraic curves defined over the real cyclotomic extension of the rationals $\mathbb{Q}(\zeta_{N}+\zeta_{-N})$, with $\zeta_{N}=e^{2\pi i/N}$. At the level of superpotential, the action of the Galois group is reproduced by a third congruence subgroup which enters in this theory, i.e. $\Gamma_{0}(N)$. An interesting implication is that the solutions of the $N$-Toda equations are related by the action of the Galois group, since the UV cusps described by the $\hat{A}_{N-1}$ models are all in the same orbit of $\Gamma_{0}(N)$.\\ Another interesting phenomenon appearing in this class of theories is that, despite the covariance of the $tt^{*}$ equations, neither the superpotential nor the ground state metric, and therefore the Berry's connection, are invariant under the action of $\Gamma(N)$. This apparent contraddiction finds a consistent explanation in the context of the abelian universal cover of the model. \\ \\ 
The paper is organized as follows: 
In section $2$ we classify up to isomorphisms all the models of the type \ref{class} with an abelian subgroup of symmetry acting transitively on the vacua. The underlyng structure of the modular curves and the relation between geometry and number theory arise naturally in the derivation. In section $3$ we provide an explicit description of these models. The target manifold is not simply connected and one needs to pull-back the model on the universal cover in order to define the Hilbert space and write the $tt^{*}$ equations. On this space the symmetry group contains also the generators of loops around the poles in the fundamental cell. The symmetry algebra is non-abelian on the universal cover and the abelian physics of the punctured plane can be recovered at the level of quantum states by considering trivial representations of the loop generators. We show that this can be done consistently with the $tt^{*}$ equations. In section $4$ we study the modular properties of these systems. First we consider the transformation of $\partial_{z}W(z)$ under the congruence subgroups and then of the superpotential. In particular we focus on the critical value of one the vacua which we use to write the $tt^{*}$ equations in the Toda form. This can be defined as holomorphic function only on the upper half plane and shifts by a costant under a transformation of $\Gamma(N)$. We connect this phenomenon to the geometry of the modular curves in the simple cases of genus $0$, i.e. with $2\leq N \leq 5$. The details of the computation of the constant for a generic $N$ are instead given in appendix. The modular transformations have also the effect of changing the basis of the symmetry generators and act on the states by modifying the representation of the symmetry group. We study this action and how the component of the ground state metric are transformed. In the last section we provide a classification of the cusps. In particular we study the behavour of the superpotentials around these points and distinguish between UV and IR critical regions. Finally we discuss the boundary conditions of the solutions and how they are related by the Galois group.

\section{Classification of the Models}

\subsection{Derivation}

Our first aim is to classify (up to isomorphisms) all the models in the class of FQHE theories \ref{class} with an abelian subgroup of symmetry acting transitively on the vacua. Since the punctured plane is not a simply connected space, one cannot define for these models a superpotential on the target manifold. So, we have to start the classification from the derivative

\begin{equation*}
 \partial_{z} W(z)= \sum_{\zeta \in \Lambda} \frac{1}{z-\zeta} + \sum_{s \in S} \frac{e(s)}{z-s},
 \end{equation*}
 
where

\begin{equation*}
\begin{split}
 & \Lambda= 2 \pi \mathbb{Z}  \oplus 2\pi \tau \mathbb{Z}, \ \tau \in \mathbb{H}, \\ \\ & e: S \longrightarrow \mathbb{C}.
\end{split} 
\end{equation*}

We stress again that the expression above is just formal and represents a meromorphic function with a simple pole at each point of $L=\Lambda \cup S$. Moreover, in this classification we allow the charges of the quasi-holes to be complex. This is mathematically consistent, since the superpotential is a complex function. \\
The action of the abelian group is transitive on the vacua and, in the case of a non trivial kernel, can always be made faithful. The transitivity implies that the set of vacua is a copy of the abelian group. In particular, given that the zeroes of $\partial_{z} W(z)$ cannot have accumulation points, it must be finitely generated. The abelian subgroups of $\mathbb{C}$ satisfying this property are lattices. Since the group acts freely also on the set of poles and the principal divisor of $\partial_{z} W(z)$ has degree $0$\footnote{This is true for a compact Riemann surface, as it turns out to be the target manifold.}, it is immediate to conclude that also $L$ is a lattice, as well as (the faithful representation of) the abelian subgroup of symmetries of our model.\\ If we want $L$ to be at least a pseudosymmetry for $\partial_{z} W(z)$, the function $e(s)$ must be extended to a multiplicative periodic character:

\begin{equation*}
\begin{split}
\partial _{z}W(z+\zeta)=e(\zeta) \partial_{z} W(z),\ \ \ \ \ \ \ \ \ \  e(\zeta+\lambda)= e(\zeta),
\end{split}
\end{equation*}

for each $\zeta \in L$, $\lambda \in \Lambda$. By definition of homomorphism, the kernel of $e$ must be a subgroup of $L$. Up to a redefinition of the initial set of holes, this is represented by the sublattice $\Lambda$, which is a symmetry for $\partial_{z} W(z)$ in a strict sense. \\ Up to this point, we have a model for each lattice $L$ and a character $e$ which is periodic of a sublattice $\Lambda \subset L$.\footnote{We are going to show that $e$ is non trivial.}\\ Given the periodicity of $e$, we can equivalently restrict the analysis to primitive characters, i.e. with trivial kernel:

\begin{equation*}
e: L/ \Lambda \longrightarrow \mathbb{C}, 
\end{equation*}

where $ L/ \Lambda \simeq \mathbb{Z}_{N_{1}} \oplus \mathbb{Z}_{N_{2}}$ for two positive integers $N_{1}, N_{2}$ such that $N_{1} \mid N_{2}$ . The fact that $e$ is primitive implies the isomorphism $ \mathbb{Z}_{N_{1}} \oplus \mathbb{Z}_{N_{2}} \simeq \mathbb{Z}_{N_{1}N_{2}}$. But, according to the chinese remainder theorem, this can be true only if the two integers are coprime. The consistency between the two conditions on the integers $N_{1},N_{2}$ requires that $N_{1}=1$ and $N_{2}=N\geq 2$, with 

\begin{equation*}
L/ \Lambda \simeq \mathbb{Z}_{N}, \ \ \ \ \ \ N \geq 2.
\end{equation*}

If we set $N=1$ we obtain the trivial case in which there are no quasi-holes. \\ In conclusion, the models are classified by couples $( E_{\Lambda},Q)$, where

\begin{itemize}
\item $E_{\Lambda}$ is the elliptic curve $ \mathbb{C}/ \Lambda$,  
\item $ Q \in E[N]= \left\lbrace P \in \mathbb{C}/ \Lambda \mid NP \in \Lambda \right\rbrace $ such that $e(Q)= e^{\frac{2\pi i}{N}}$.
\end{itemize}

The set $E[N]$ is called the N-torsion subgroup of the additive torus group $\mathbb{C}/ \Lambda$. \\ Once $\Lambda$ and the level $N$ are fixed, the choice of the torsion point specifies an embedding of the cyclic subgroup $L/ \Lambda \simeq \mathbb{Z}_{N}$ in the elliptic curve. In particular, the torsion point must be of order $N$, i.e. such that $NQ \in \Lambda$ but $nQ \not \in \Lambda$ for $ 1<n<N$.

\subsection{Modular Curves}\label{modular}

Since we are classifying models up to strict equivalence, we have to identify those which are related by an isomorphism.  It is known that there is a bijection between the set of equivalence classes of elliptic curves endowned with a N-torsion point and the space 

\begin{equation*}
Y_{1}(N)= \mathbb{H}/\Gamma_{1}(N),
\end{equation*}

where $\Gamma_{1}(N)$ is a subgroup of $SL(2,\mathbb{Z})$ defined by the congruence condition

\begin{equation*}
\Gamma_{1}(N)= \left\lbrace \gamma \in SL(2,\mathbb{Z}) : \gamma=\begin{pmatrix} a & b \\ c & d

\end{pmatrix}= \begin{pmatrix} 1 & * \\ 0 & 1 \end{pmatrix} \mathrm{mod} \ N \right\rbrace .
\end{equation*}

A complete proof of this result can be found in \cite{rif2}. It is worth to recall that a matrix $\gamma \in SL(2,\mathbb{Z})$ acts on the upper half plane by the usual fractional linear transformation

\begin{equation}
\tau^{\prime}=\begin{pmatrix} a & b \\ c & d \end{pmatrix}\tau=  \frac{a\tau+b}{c\tau+d},
\end{equation}

and induces on the points of the elliptic curve the isogeny $z + \Lambda_{\tau} \longrightarrow m z + \Lambda_{\tau^{\prime}} $, for some $m \in \mathbb{C}$ such that $m\Lambda_{\tau}=\Lambda_{\tau^{\prime}}$. These maps are the only bijections which preserve the group structure of the elliptic curve. With these definitions it is immediate to show that the enhanced elliptic curve $(E_{\Lambda_{\tau^{\prime}}},Q)$ is isogenous to $\left( E_{\Lambda_{\tau}}, 2\pi/N +\Lambda_{\tau}\right)$, where $\tau^{\prime}=\gamma(\tau)$ for some $\gamma \in SL(2,\mathbb{Z})$, and that transformations of $\Gamma_{1}(N)$ are the only ones which preserve the choice of the torsion point. So, we can define the moduli space for $\Gamma_{1}(N)$ as

\begin{equation*}
 S_{1}(N)= \lbrace \left( E_{\Lambda_{\tau}}, 2\pi/N +\Lambda_{\tau}\right) , \tau \in \mathbb{H}  \rbrace / \sim 
\end{equation*}

where $\tau \sim \tau^{\prime}$ if and only if $\Gamma_{1}(N)\tau=\Gamma_{1}(N)\tau^{\prime}$, and state the bijection 

\begin{equation*}
 S_{1}(N) \overset{\sim}{\longrightarrow} Y_{1}(N).
\end{equation*}

The space $Y_{1}(N)$ is topologically a complex manifold with cusps and can be compactified. One first has to extend the action of $SL(2,\mathbb{Z})$ to the rational projective line $\mathbb{P}^{1}(\mathbb{Q})= \mathbb{Q} \cup \lbrace \infty  \rbrace$. Given  $\begin{pmatrix} p & q \\ r & t \end{pmatrix} \in SL(2,\mathbb{Z})$, we have 

\begin{equation*}
\begin{split}
& \mathbb{H} \longrightarrow \mathbb{\overline{H}}= \mathbb{H} \cup \mathbb{P}^{1}(\mathbb{Q}),  \\ \\ 
 \begin{pmatrix} p & q \\ r & t \end{pmatrix} \frac{a}{c}=&  \frac{pa+qc}{ra+tc},  \hspace{2cm } \begin{pmatrix} p & q \\ r & t \end{pmatrix} \infty = \frac{p}{q}
\end{split}
\end{equation*}

with $a,c \in \mathbb{Z}$ such that $\mathrm{gcd}(a,c)=1$.\\ The set of cusps is given by $C_{\Gamma_{1}(N)}=\mathbb{P}^{1}(\mathbb{Q})/ \Gamma_{1}(N)$ and is finite. By adding these points to $Y_{1}(N)$ one obtains

\begin{equation*}
Y_{1}(N)\longrightarrow X_{1}(N)= \overline{\mathbb{H}}/\Gamma_{1}(N).
\end{equation*}

The space $X_{1}(N)$ is called modular curve for $\Gamma_{1}(N)$ and can be shown to have the structure of a compact Riemann surface. Since the two dimensional lattice becomes degenerate when $\tau$ approaches the rational projective line, the cusps cannot be strictly considered as members of this class of theories, but rather as critical limits\footnote{It will be clear later that the cusps can be interpreted as RG flow fixed points.}.\\ So, we learn that the whole space of models can be written as union of connected components labelled by the integer $N$:

\begin{equation*}
\mathcal{A}= \bigcup _{N \geq 2} X_{1}(N),
\end{equation*}

where $X_{1}(N)$ parametrizes the subclass of theories of level $N$. It follows from the derivation that each point on $\mathcal{A}$ identifies a model up to isomorphisms. \\ The next step is to classify the vacua for this family of theories. Let us consider the modular curve of level $N$. The derivative of the superpotential is an elliptic function on the elliptic curve $\mathbb{C}/ \Lambda_{\tau}$ which has a simple pole at each point $kQ + \Lambda_{\tau},\  k=0,...,N-1$ and a simple zero at each point $P+kQ+ \Lambda_{\tau}, \ k=0,...,N-1$ for some $P \in C/ \Lambda_{\tau}$ such that $\partial_{z} W(P)=0$. Given that the principal divisor is vanishing by Abel's theorem, we get

\begin{equation*}
\mathrm{div}(\partial_{z} W(z))= \sum_{0\leq k < N} ([P]+k[Q])-\sum_{0\leq k < N} (k[Q])= N[P]=0.
\end{equation*}

So we deduce that also $P$ must be a torsion point and, once known, we can construct the whole set of vacua with the action of the symmetry generators. In order to find $P$, one can define the Weil pairing of order $N$ between torsion points:

\begin{equation*}
\begin{split}
& e_{N}: E[N] \ \mathrm{x} \ E[N]\longrightarrow \mu_{N}, \ \ \ \ \ \ \ \ \ \ \mu_{N}=\lbrace z \in \mathbb{C}\mid z^{N}=1\rbrace, \\ \\ \ \ \ \ & \hspace{3cm}e_{N}(P,Q)= \frac{F_{P}(z+Q)}{F_{P}(z)},
\end{split}
\end{equation*}

where $F_{P}(z)$ is any elliptic function with simple poles at $kQ+\Lambda_{\tau}$ and simple zeroes at $P+kQ+\Lambda_{\tau}$. Since $F_{P}(z)$ is unique up to multiplicative constants, there is no ambiguity in the definition. The Weil pairing is a sort of inner product on $E[N]$ and can be shown to be alternating, bilinear and non degenerate \cite{rif2,rif3}. These properties imply that, once the torsion point of the poles $Q$ is picked, a torsion point of the vacua $P$ is coupled by the condition $e_{N}(P,Q)= e^{\frac{2\pi i}{N}}$. Moreover, another point $P^{\prime}$ which has the same pairing with $Q$ must be of the form $P^{\prime}=P+kQ$ for some integer $k$. We can also give an explicit expression for $e_{N}$ \cite{rif2}. Chosen $\left( 2\pi/N + \Lambda_{\tau}, 2\pi \tau /N+ \Lambda_{\tau}\right) $ as basis of generators for the torsion group, the formula for the Weil pairing of $P,Q \in E[N]$ is 

\begin{equation*}
e_{N}(P,Q)= e^{2\pi i \ \mathrm{ det} \gamma /N},
\end{equation*}

where 

\begin{equation*}
\begin{bmatrix}
P \\ Q
\end{bmatrix}= \gamma \begin{bmatrix}
2\pi/N + \Lambda_{\tau} \\ 2\pi \tau/N + \Lambda_{\tau}
\end{bmatrix}, \ \ \ \ \ \mathrm{for} \ \gamma \in M_{2}(\mathbb{Z}).
\end{equation*}

This definition is actually indipendent of how the basis of the torsion group is chosen. One can see that, if $P$ and $Q$ generate $E[N]$, the matrix $\gamma$ is invertible and $e_{N}(P,Q)$ is a primitive complex $N$th root of unity. Moreover, such expression is preserved under isogenies.\\
Now we can classify models and vacua with triplets  $(E_{\Lambda_{\tau}}, P, Q)$ which denote elliptic curves with associated two torsion data. If we set $Q=2\pi/N +\Lambda_{\tau}$ with a modular transformation, a torsion point satisfying the Weil condition is $P=-2\pi \tau/N+\Lambda_{\tau}$. It is straightforward to see that the congruence subgroup of $SL(2,\mathbb{Z})$ which preserves both the torsion points is:

\begin{equation*}
\Gamma(N)= \left\lbrace \gamma \in SL(2,\mathbb{Z}) : \gamma= \begin{pmatrix}
a & b \\ c & d \end{pmatrix} = \begin{pmatrix} 1 & 0 \\ 0 & 1 \end{pmatrix} \mathrm{mod} \ N\right\rbrace,
\end{equation*}

also called principal congruence subgroup. Similarly to the case of $S_{1}(N)$, one shows that the moduli space for the enhanced elliptic curves of principal type is 

\begin{equation*}
 S(N)= \lbrace \left( E_{\Lambda_{\tau}}, -2\pi\tau/N+\Lambda_{\tau}, 2\pi/N +\Lambda_{\tau}\right) , \tau \in \mathbb{H}  \rbrace / \sim 
\end{equation*}

where $\tau \sim \tau^{\prime}$ if and only $\Gamma(N)\tau=\Gamma(N)\tau^{\prime}$. This set of equivalence classes is therefore isomorphic to the complex manifold

\begin{equation*}
Y(N)= \mathbb{H}/ \Gamma(N).
\end{equation*}

As for $Y_{1}(N)$, we can compactify such space by adding the set of cusps $C_{\Gamma(N)}= \mathbb{P}^{1}(\mathbb{Q})/\Gamma(N)$ and obtain the compact Riemann surface

\begin{equation*}
Y(N)\longrightarrow X(N)= \overline{\mathbb{H}}/\Gamma(N),
\end{equation*}

which is known as the N-modular curve of principal type.\\ This curve represents the spectral cover of the space of models of level $N$ \cite{rif12}. Given that $\Gamma(N)$ is a normal subgroup of $\Gamma_{1}(N)$ with coset group $\Gamma_{1}(N)/\Gamma(N)\simeq \mathbb{Z}_{N}$, the degree of the cover is $[\Gamma_{1}(N):\Gamma(N)]=N$, which is exactly the number of vacua up to periodic identification.\\ A series of results which allows to count the cusps and describe the set $C_{\Gamma(N)}$ can be found in \cite{rif2}. In particular, denoting with $\begin{bmatrix} a \\ c \end{bmatrix}, \begin{bmatrix} a^{\prime} \\ c^{\prime} \end{bmatrix}$ two vectors of $\mathbb{Z}^{2}$, one can show that

\begin{equation}\label{width1}    
\begin{bmatrix} a^{\prime} \\ c^{\prime} \end{bmatrix}= \gamma \begin{bmatrix} a \\ c \end{bmatrix}, \ \mathrm{for} \ \gamma \in \Gamma(N) \Longleftrightarrow \begin{bmatrix} a^{\prime} \\ c^{\prime} \end{bmatrix} =  
\begin{bmatrix} a \\ c \end{bmatrix} \ \mathrm{mod} \ N.
\end{equation}

From this result, letting $s=a/c$ and $s^{\prime}= a^{\prime}/c^{\prime}$ two elements of $\mathbb{Q}\cup\left\lbrace \infty\right\rbrace $ such that $\mathrm{gcd}(a,c)= \mathrm{gcd}(a^{\prime},c^{\prime})=1$, it follows the equivalence relation

\begin{equation}\label{width2}   
\Gamma(N)s^{\prime}= \Gamma(N)s \Longleftrightarrow \begin{bmatrix} a^{\prime} \\ c^{\prime} \end{bmatrix} = \pm 
\begin{bmatrix} a \\ c \end{bmatrix} \ \mathrm{mod} \ N,
\end{equation}

where the sign `$-$'  keeps into account the projective action of the modular group on $\mathbb{P}^{1}(\mathbb{Q})$. Once collected the rationals in $\Gamma(N)$-classes according to this theorem, one can go futher and find also the cusps of $\Gamma_{1}(N)$. It is enough for this purpose to identify the cusps of $\Gamma(N)$ which belong to the same orbit of $T:\tau\rightarrow \tau +1$, i.e. the generator of $\Gamma_{1}(N)/\Gamma(N)$. Indeed, these points represent vacua of the same theory and are projected on the same cusp of $X_{1}(N)$.\\ From a geometrical point of view the cusps are the points for which the covering map $X(N)\rightarrow X_{1}(N)$ degenerates and the dimension of the orbit of $T$ is generically less than $N$. In physical terms, these represent the fixed points of the RG flow in the space of couplings. Being more precise, the cusps of $\Gamma(N)$ which are stabilized by $T$ are expected to be UV fixed points of the RG flow. Indeed, in this limits the zeroes of $\partial_{z}W(z)$ tend to a unique vacuum of order $N$ and the BPS states of the theory become consequently massless, implying that we are approaching a conformal field theory \cite{rif10}. On the other hand, in the IR limits the vacua become infinitely separated from each other and decouple at the leading order. If they are not simple zeroes of $\partial_{z}W(z)$, i.e. the corresponding orbit of $T$ has dimension between $1$ and $N$, these cusps represent again conformal fixed points on the spectral cover. Instead, if the vacuum is a simple zero and the orbit of the cusp under $T$ has dimension exactly $N$, we are dealing with a free massive field theory. We will provide a more precise description of the physics around the cusps in section $5$.

\subsection{The Role of Number Theory}

Another congruence subgroup of $SL(2,\mathbb{Z})$ which plays an important role in this classification is 

\begin{equation*}
\Gamma_{0}(N)= \left\lbrace \gamma \in SL(2,\mathbb{Z}): \gamma= \begin{pmatrix}
a & b \\ c & d \end{pmatrix} = \begin{pmatrix} * & * \\ 0 & * \end{pmatrix} \mathrm{mod} \ N \right\rbrace.
\end{equation*}

The three congruence subgroups that we have defined satisfy the chain of inclusions $\Gamma(N)\subset \Gamma_{1}(N) \subset \Gamma_{0}(N) \subset SL(2,\mathbb{Z})$. Moreover, $\Gamma_{1}(N)$ is a normal subgroup of $\Gamma_{0}(N)$. The action of $\Gamma_{0}(N)$ on the space of models is more clear when we choose $Q= \frac{2\pi}{N}$. It is evident that $\gamma \in \Gamma_{0}(N)$ does not leave the model invariant and maps the torsion point into an inequivalent one $\gamma^{*}Q= 2\pi \frac{(c\tau+d)}{N} \sim \frac{2\pi d}{N}$, for an integer $d$ coprime with $N$. The effect on the model corresponds to modify the character by the formula $\sigma_{a}(e(Q))= e(Q)^{a}=e^{\frac{2\pi i a}{N}}$, where $a$ is the inverse of $d$ in $\mathbb{Z}_{N}$. Therefore, it is manifest that $\Gamma_{0}(N)$ reproduces the action of the Galois group of the cyclotomic extension $[\mathbb{Q}(\zeta_{N}):\mathbb{Q}]$. This number field is obtained by adjoining a primitive $N$-th root of unity $\zeta_{N}$ to the rational numbers. Such remarkable connection reveals the algebraic nature of the modular curves. The above formula suggests to define a character $e_{l}(Q)=e^{\frac{2\pi i l}{N}}$ depending on an integer $l$ coprime with $N$, which we call co-level of the modular curve. It is clear that, for a fixed lattice $\Lambda_{\tau}$, a different choice of the co-level corresponds to pick a different point on the space of models. With this definition, a transformation of $\Gamma_{0}(N)$ can be seen as a permutation of the co-levels.\\ We can be more precise about these statements by keeping into account that $\Gamma_{0}(N)$ contains the matrix $-I$. We know that a point on $X_{1}(N)$ parametrizes an elliptic curve and a specific embedding of $\mathbb{Z}_{N}$. However, a cyclic group has always two generators which are one the inverse of the other. It is clear that $-I$ acts on the model as a parity transformation, since it does not change the point on $X_{1}(N)$ but inverts the sign of the torsion point. This means that the co-levels $l$ and $N-l$ are actually two descriptions of the same model. So, except for the trivial case of $N=2$, since $-I$ is in $ \Gamma_{0}(N)$ but not in $\Gamma_{1}(N)$ the degree of the cover $\mathbb{H}/ \Gamma_{0}(N)\rightarrow \mathbb{H}/ \Gamma_{1}(N)$ is $\left[ \Gamma_{0}(N):\Gamma_{1}(N) \right]/2  = \phi(N)/2$,  where $\phi(N)$ is the Euler totient function which count the elements of $\left\lbrace 0,...,N-1\right\rbrace$ coprime with $N$. As one can expect, this is also the degree of the number field defining this class of theories, which is actually the real cyclotomic extension $\mathbb{Q}(\zeta_{N}+\zeta_{-N})$. The cusps of $\Gamma_{1}(N)$ fall in equivalence classes of the Galois group described by the set $C_{\Gamma_{0}(N)}= \mathbb{P}^{1}(\mathbb{Q})/\Gamma_{0}(N)$. We will discuss the orbits of $\Gamma_{0}(N)$ in the set of critical theories in section $5$.

\section{Geometry of the Models}

\subsection{The Model on the Target Manifold}

Now we want to translate our abstract classification into an explicit description of these models. A fundamental property of the elliptic functions is that they are uniquely specified (up to multiplicative constants) by the positions and orders of their zeroes and poles. Let us focus on the modular curve of level $N$ and co-level $l$ and choose $(P,Q)=(-\frac{2\pi l \tau}{N}+\Lambda_{\tau},\frac{2\pi }{N}+\Lambda_{\tau})$ as torsion points. For convenience we invert the lattices of poles and vacua with respect to the previous derivation. This can be done with the translation $z \rightarrow z- \frac{2\pi l \tau}{N}$. The derivative of the superpotential for this class of theories is

\begin{equation}
\partial_{z} W^{(N,l)}(z;\tau)= \sum_{k=0}^{N-1} e^{\frac{2\pi i l k}{N}} \left[ \zeta \left( z-\frac{2\pi}{N}(l\tau + k);\tau\right) +\frac{2 \eta_{1}k}{N}\right], 
\end{equation}

where  $  z \in \mathcal{S}= \mathbb{C} \setminus \left\lbrace \frac{2\pi}{N}(l\tau+k)+\Lambda_{\tau}, \ k=0,....,N-1; \ \Lambda_{\tau}= 2\pi\mathbb{Z} \oplus 2\pi \tau \mathbb{Z}\right\rbrace $. \\ The Weierstrass zeta function $\zeta(z;\tau)$ is defined with the conventions $\eta_{1}=\zeta(\pi;\tau)$, $ \eta_{2}=\zeta(\pi \tau;\tau)$ \cite{rif6}. This function is meromorphic on $\mathcal{S}$ with simple poles and simple zeroes respectively in $\frac{2\pi(l\tau+k)}{N}+\Lambda_{\tau}$ and $\frac{2\pi k}{N}+\Lambda_{\tau}$, $k=0,...,N-1$. By definition of elliptic function, the sum of the residue inside the fundamental cell is vanishing. From a physical point of view this means that the flux of the magnetic field is cancelled by that of the quasi-holes charges. Therefore, even though with beatiful mathematical properties, this class of theories is not interesting for the FQHE phenomenology.\\ The algebra of the abelian symmetry group of this model is generated by three operators $\sigma,A,B$, defined by the actions

\begin{equation}\label{transf}
\begin{split}
& \sigma: z\longrightarrow z+ 2\pi/N ;\hspace{2cm} \partial_{z} W^{(N,l)}\left( z+2\pi/N;\tau\right) = e^{\frac{2\pi i l}{N}}\partial _{z}W^{(N,l)}\left( z;\tau\right),  \\ \\ 
& A: z\longrightarrow z+2\pi; \hspace{2.4cm} \partial_{z} W^{(N,l)}\left( z+2\pi;\tau\right) = \partial_{z} W^{(N,l)}\left( z;\tau\right),  \\ \\
& B: z\longrightarrow z +2\pi \tau ; \hspace{2.1cm} \partial_{z} W^{(N,l)}\left( z+2\pi \tau;\tau\right) = \partial_{z} W^{(N,l)}\left( z;\tau\right) ,
\end{split}
\end{equation}

with the additional relation $ \sigma^{N}= A$. These transformations follows from the quasi-periodicity properties of the Weierstrass function. The double periodicity of $\partial _{z}W^{(N,l)}\left( z;\tau\right) $ allows to identity points wich differ by an element of $\Lambda_{\tau}$, so that the model is naturally projected on the torus $\mathcal{K}=\mathcal{S}/\Lambda_{\tau} $. With this identification we can work with just $N$ critical points, denoted by the equivalence classes $\left[ \frac{2\pi k}{N}\right], k=0,...,N-1 $. From the property $\zeta(-z;\tau)=-\zeta(z;\tau)$, we find that the superpotentials of co-levels $l$ and $N-l$ are actually related by a parity transformation

\begin{equation}\label{colevel}
\partial_{z} W^{(N,N-l)}(-z;\tau)=-\partial_{z} W^{(N,l)}(z;\tau)
\end{equation}

and therefore, as pointed out in the previous section, they describe the same model with inverse torsion points generating the $\mathbb{Z}_{N}$ symmetry.\\ We can also write for this class of theories a symbolic `primitive'

\begin{equation}\label{symb}
W^{(N,l)}(z;\tau)= ``\sum_{k=0}^{N-1} e^{\frac{2\pi i l k}{N}} \log  \Theta \begin{bmatrix} \frac{1}{2}-\frac{l}{N} \\ \frac{1}{2}-\frac{k}{N}\end{bmatrix}\left( \frac{z}{2\pi};\tau \right) ",
\end{equation}

where $\Theta\begin{bmatrix} \alpha \\ \beta \end{bmatrix}(z;\tau)$ is the theta function of character $(\alpha,\beta) \in \mathbb{R}$ and quasi-periods $1,\tau$. This function is multi-valued and cannot be really considered as superpotential of the model. Indeed, the target manifold is not simply connected and we cannot find a true primitive on this space. 

\subsection{Abelian Universal Cover}

In order to define a superpotential, as well as the Hilbert space of the theory, we need to pull-back the model on the universal cover of $\mathcal{S}$. We consider the smallest one, i.e. the abelian universal cover. This can be defined in an abstract way as the space of curves 

\begin{equation}\label{def}
\begin{split}
& \hspace{3cm}\mathcal{H}= \left\lbrace p:[0,1]\longrightarrow  \mathcal{S}, p(0)=p^{*} \in \mathcal{S} \right\rbrace / \sim, \\ \\ 
& \sim: \hspace{2cm} 
p\sim q=
\begin{cases}
p(1)=q(1), \\ 
p \cdot q^{-1} =0 \ \mathrm{in \ the \ homology \ group} \  H_{1}(\mathcal{S},\mathbb{Z}).
\end{cases}
\end{split}
\end{equation}

We can pull-back $ \partial_{z} W^{(N,l)}(z;\tau)$ on this space and give a formal definition of superpotential:

\begin{equation}
W^{(N,l)}(p;\tau)= \int_{p} \partial_{z} W^{(N,l)}(z;\tau) dz 
\end{equation}

\begin{figure}[!h]
\centering
\begin{tikzpicture}
\coordinate [label=below left:$\gamma_{a}$] (A)  at (7,0);
\coordinate [label=above left:$\gamma_{b}$] (B)  at (1,1.5);
\coordinate [label=below left: $z$] (Z) at (0,0);
\coordinate[label= x] (x) at (1.5,1);
\coordinate[label= x ] (x) at (3,1);
\coordinate[label= x ] (x) at (6.8,1);
\coordinate[label= x ] (x) at (11,1);
\coordinate[label= x ] (x) at (13,1);
\coordinate[label=  . . . . . . . . . .] (x) at (4.87,1);
\coordinate[label=   . . . . . . . . . . . ] (x) at (8.88,1);
\coordinate [label= above: $\ell_{1} $]    (L)     at (2,2);
\coordinate [label= above: $\ell_{2} $]    (L)     at (4,2);
\coordinate [label= above: $\ell_{k} $]    (L)     at (7.5,1.7);
\coordinate [label= above: $\ell_{N-1} $]    (L)     at (11.5,1.47);

\draw  [->](0,0)--(13,0);
\draw  (13,0)--(15,3);
\draw [->](0,0)--(2,3);
\draw  (2,3)--(15,3);
\draw  ((0,0) to [out=35, in=180] (2,2);
\draw [->](0,0) to [out=0, in=0] (2,2);
\draw  ((0,0) to [out=2, in=180] (4,2);
\draw [->](0,0) to [out=0, in=0] (4,2);
\draw  ((0,0) to [out=0, in=180] (7.5,1.7);
\draw [->](0,0) to [out=0, in=0] (7.5,1.7);
\draw  ((0,0) to [out=0, in=180] (11.5,1.47);
\draw [->](0,0) to [out=0, in=0] (11.5,1.47);

\end{tikzpicture}
\caption{Homology generators}\label{homology}
\end{figure}
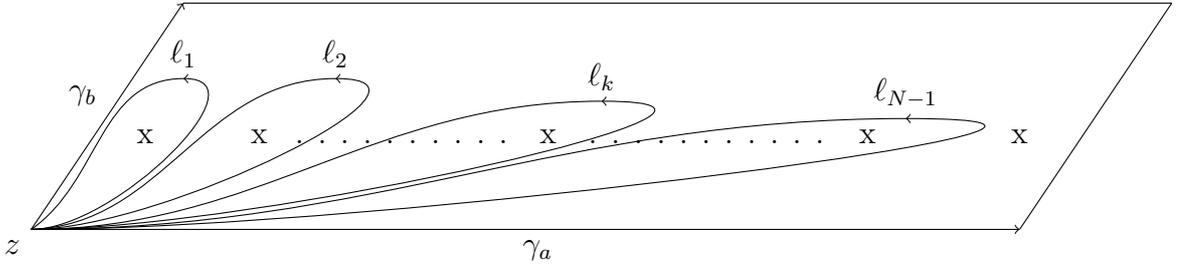

Contrary to \ref{symb}, this is a well defined function which assigns a single value to each equivalence class of paths in $\mathcal{H}$.\\
In the covering model we have many more vacua than before, but at the same time a larger symmetry group to classify them. An evident subgroup is the deck group of the cover, i.e. the homology. Let us fix a point $z \in \mathcal{S}$. We choose as local basis of $H_{1}(\mathcal{S};\mathbb{Z})$ the one given by the curves $\gamma_{a},\gamma_{b}$, obtained by applying respectively the lattice vectors $a=2\pi,b=2\pi \tau$ to the base point $z$, and the anticlockwise loops $\ell_{0},...,\ell_{N-2}$ encircling the first $N-1$ poles contained in the cell with sides $\gamma_{a},\gamma_{b}$ (Figure \ref{homology}). The action on the superpotential of the corresponding operators is

\begin{equation}\label{integrals}
\begin{split}
& A^{*}W^{(N,l)}(p;\tau)= \int_{\gamma_{a}\cdot p} \partial_{z} W^{(N,l)}(z;\tau)dz = W^{(N,l)}(p;\tau) + \int_{\gamma_{a}} \partial_{z} W^{(N,l)}(z;\tau)dz, \\ \\ 
& B^{*}W^{(N,l)}(p;\tau)= \int_{\gamma_{b}\cdot p} \partial_{z} W^{(N,l)}(z;\tau)dz = W^{(N,l)}(p;\tau) + \int_{\gamma_{b}} \partial_{z} W^{(N,l)}(z;\tau)dz, \\ \\ 
& L_{k}^{*}W^{(N,l)}(p;\tau)=  \int_{\ell _{k}\cdot p} \partial_{z} W^{(N,l)}(z;\tau)dz = W^{(N,l)}(p;\tau) + \oint_{\ell_{k}} \partial_{z} W^{(N,l)}(z;\tau)dz.
\end{split}
\end{equation}

Compatibly with the definition of symmetry, the superpotential shifts by a constant under the action of the homology generators. \\ \\ 

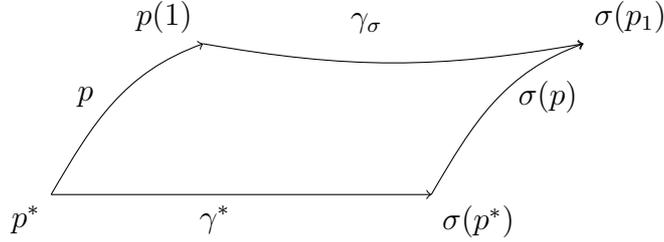
\begin{figure}[h!]
\centering
\begin{tikzpicture}
\coordinate [label=below left:$\gamma^{*}$] (A)  at (2.5,0);
\coordinate [label=above left:$\gamma_{\sigma}$] (B)  at (4.5,2);
\coordinate [label=below left: $p^{*}$] (P) at (0,0);
\coordinate [label=right:$\sigma (p)$] (A)  at (6,1.3);
\coordinate [label=left:$p$] (A)  at (0.7,1.3);
\coordinate [label=below right:$ \sigma(p^{*})$] (A)  at (5,0);
\coordinate [label=above left:$p(1)$] (A)  at (2,2);
\coordinate [label=above right:$\sigma(p_{1})$] (A)  at (7,2);

\draw  [->](0,0)--(5,0);
\draw  [->] ((0,0) to [out=60, in=200] (2,2);
\draw [->](2,2) to [out=-10, in=190] (7,2);
\draw  [->] ((5,0) to [out=60, in=200] (7,2);

\end{tikzpicture}
\caption{Definition of $\gamma_{\sigma}$}\label{sigma}
\end{figure}

In order to define the action of $\sigma$ on $\mathcal{H}$ we need to specify a path $\gamma_{\sigma}$ connecting $z$ to $z+\frac{2\pi}{N}$. Let us fix the point $p^{*}$ in the definition \ref{def}. We choose a path $\gamma^{*}$ connecting $p^{*}$ to $p^{*}+\frac{2\pi}{N}$. If the point is not orizontally aligned with the poles we can take for instance a straight line. Now, given a curve $p$ in $\mathcal{H}$, we define $\gamma_{\sigma}$ such that $\gamma_{\sigma} \cdot p \ $= $ \sigma (p) \cdot \gamma^{*}$ in $H_{1}(\mathcal{S};\mathbb{Z})$, where $\sigma(p)$ denotes the curve $p$ shifted by $\frac{2\pi}{N}$ (Figure \ref{sigma}). Choosing the integration constant so that $W^{(N,l)}(\gamma^{*};\tau)=0$, we get the transformation of the superpotential

\begin{equation}\label{exact}
\begin{split}
\sigma^{*}W^{(N,l)}(p;\tau) = & \int_{\gamma_{\sigma}\cdot p} \partial_{z} W^{(N,l)}(z;\tau)dz=
\int_{\sigma (p)\cdot \gamma_{*}} \partial_{z} W^{(N,l)}(z;\tau)dz= \\ \\ & e^{\frac{2\pi i l}{N}} W^{(N,l)}(p;\tau) + W^{(N,l)}(\gamma^{*};\tau)= e^{\frac{2\pi i l}{N}} W^{(N,l)}(p;\tau) . 
\end{split}
\end{equation}

Inequivalent definitions of $\gamma_{\sigma}$ differ by compositions with the loops $\ell_{k} $ and generate a different constant $W^{(N,l)}(\gamma^{*};\tau)$ in the transformation. Indipendently from the choice, by composing $N$ times $\gamma_{\sigma}$ we obtain an curve in $H_{1}(\mathcal{S};\mathbb{Z})$ homologous to $\gamma_{a}$, consistently with the operatorial relation $\sigma^{N}=A$. \\ \\ 

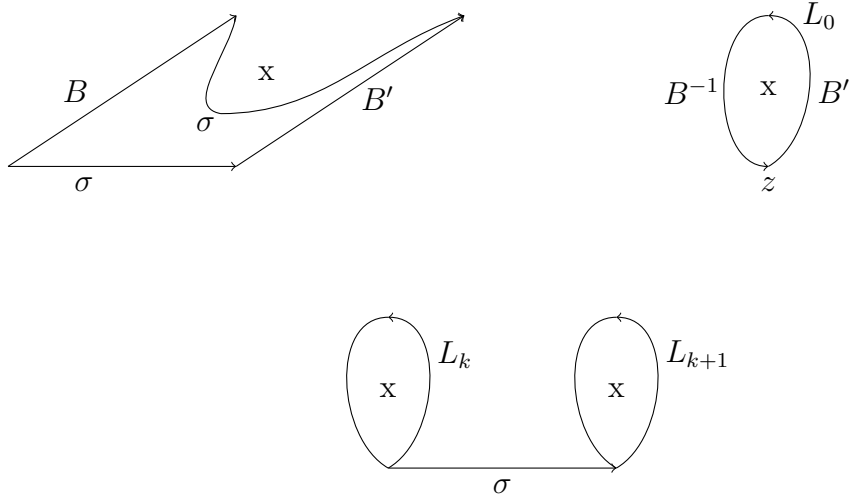
\begin{figure}[h!]
\centering
\begin{tikzpicture}
\coordinate [label=left:$B$] (A)  at (1.2,1);
\coordinate [label=right:$B^{\prime}$] (B)  at (4.5,0.9);
\coordinate [label=left:$B^{-1}$] (A)  at (9.5,1);
\coordinate [label=right:$B^{\prime}$] (B)  at (10.5,1);
\coordinate [label= x] (X) at (3.4,1);
\coordinate [label= x] (X) at (10,0.8);
\coordinate [label= x] (X) at (5,-3.2);
\coordinate [label= x] (X) at (8,-3.2);
\coordinate [label= below:$\sigma$] (s) at (2.6,0.8);
\coordinate [label= below:$z$] (X) at (10,0);
\coordinate [label= below:$\sigma$] (s) at (1,0);
\coordinate [label= below:$\sigma$] (s) at (6.5,-4);
\coordinate [label= right:$L_{k}$] (s) at (5.5,-2.5);
\coordinate [label= right:$L_{k+1}$] (s) at (8.5,-2.5);
\coordinate [label= right:$L_{0}$] (s) at (10.3,2);

\draw  [->](0,0)--(3,0);
\draw  [->](0,0)--(3,2);
\draw  [->](3,0)--(6,2);
\draw  [->](5,-4)--(8,-4);
\draw  ((3,2) to [out=-100, in=170] (2.8,0.7);
\draw  [->] ((2.8,0.7) to [out=00, in=200] (6,2);
\draw  [->] ((10,2) to [out=180, in=180] (10,0);
\draw  [->] ((10,0) to [out=30, in=00] (10,2);
\draw   ((5,-4) to [out=150, in=180] (5,-2);
\draw  [->] ((5,-4) to [out=30, in=00] (5,-2);
\draw   ((8,-4) to [out=150, in=180] (8,-2);
\draw  [->] ((8,-4) to [out=30, in=00] (8,-2);

\end{tikzpicture}
\caption{Algebra of curves}\label{algebr}
\end{figure}

The generators $\sigma, A,B,(L_{k})_{k=0,...,N-2}$ represent a complete basis for the symmetry group of the covering model. It is clear from Figure \ref{algebr} that the algebra is non-abelian on the space of curves. Indeed, the generator $\sigma$ do not commute with $B$ and $L_{k}$ according to the algebraic relations

\begin{equation}\label{identity}    
\sigma B= L_{0} B \sigma, \hspace{2cm} \sigma L_{k}= L_{k+1} \sigma, \ \ \ k=0,...,N-2,
\end{equation}

where we have $L_{N-1}= \prod\limits_{k=0}^{N-2}L_{k}^{-1}$. We see that the obstruction to abelianity on the universal cover is represented by the generators of loops. In the faithful representation on the plane these act trivially and $A,B,\sigma$ commute. This is not inconsistent, since the physics must be abelian on the target manifold but not necessarily on the covering model. The abelianity which we have asked in the classification can be recoverd at the quantum level by projecting the Hilbert space on the trivial representation of $L_{k}$. We will discuss this point in the next section. There are not problems instead in the commutation between $\sigma$ and $A$ and clearly the subgroup given by the homology is abelian. 

\newpage

Now we can classify the vacua of the model defined on $\mathcal{H}$. The vacua on the target manifold are represented by the points $z_{k}= \frac{2\pi k}{N}, k=0,...,N-1,$ up to periodic identification of $\Lambda_{\tau}$.
For these points there is an ambiguity in the definition of the cycle $\gamma_{b}$, since there is a pole along the side of the fundamental cell. We choose conventionally the homology path which bypasses the pole on the left. Now that the homology based on the critical points is well defined, we fix $p^{*}$ away from the set of vacua and pick a curve $p_{0}$ connecting $p^{*}$ to $z_{0}$ as representative of $z_{0}$ in $\mathcal{H}$.  All the other representatives in the corresponding fiber can be obtained by composing $p_{0}$ with the cycles of $H_{1}(\mathcal{S},\mathbb{Z})$ based on $z_{0}$. Moreover, we can map fibers over different points to each other with the action of $\sigma$. Therefore, the vacua on $\mathcal{H}$ are labelled by the curves

\begin{equation}\label{vacua}
p_{k;m,n,n_{0},...,n_{N-2}}= \gamma_{a}^{m} \cdot \gamma_{b}^{n} \cdot \prod_{r=0}^{N-2} \ell_{r}^{n_{r}} \cdot \gamma_{\sigma}^{k} \cdot p_{0}
\end{equation}

where $k=0,...,N-1,$ and $m,n,n_{0},...,n_{N-2} \ \in \mathbb{Z}$.  By definition of abelian universal cover, each fiber is a copy of the homology group.\\ We can also provide the corresponding critical values by computing the integrals in  \ref{integrals}. Let us take again the curve $p_{0}$. By the relation $A=\sigma^{N}$ and the residue formula we have 

\begin{equation}
\begin{split}
& A^{*}W^{(N,l)}(p_{0};\tau)= W^{(N,l)}(p_{0};\tau),\\ \\ 
&  L_{k}^{*}W^{(N,l)}(p_{0};\tau)= W^{(N,l)}(p_{0};\tau) + 2\pi i \ e^{\frac{2\pi i l k}{N}}.
 \end{split}
\end{equation}

Moreover, from the first algebraic identity in \ref{identity}, we easily get

\begin{equation}
B^{*}W^{(N,l)}(p_{0};\tau)= W^{(N,l)}(p_{0};\tau) + \frac{2 \pi i}{e^{\frac{2\pi i l}{N}}-1}.
\end{equation}

Finally, by acting with the generators $\sigma, A,B,(L_{k})_{k=0,...,N-2}$ on $W^{(N,l)}(p_{0};\tau)$, we obtain the whole set of critical values

\begin{equation}\label{critical}
\begin{split}
W^{(N,l)} & (p_{k;m,n,n_{0},...,n_{N-2}};\tau)  = A^{* m} B^{*  n} \prod_{0\leq r \leq N-2} L_{r}^{*  n_{r}} \sigma^{*  k} \ W^{(N,l)}(p_{0};\tau)= \\ \\
  &  e^{ \frac{2 \pi i l k}{N}} W^{(N,l)}(p_{0};\tau)+ n \frac{2 \pi i}{e^{\frac{2\pi i l}{N}}-1} + \sum_{r=0}^{N-2} n_{r} \ 2\pi i \ e^{\frac{2\pi i l r}{N}} .
  \end{split}
\end{equation}

\subsection{$tt^{*}$ Geometry}

\subsubsection{$tt^{*}$ Geometry in $2$ Dimensions}

In this section we are going to study the geometry of vacua of this class of models. We begin by recalling some known results about $tt^{*}$ geometry in two dimensions \cite{rif4}. In $\mathcal{N}=(2,2)$ supersymmetric Landau-Ginzburg theories in two dimensions we can consider two types of deformations, involving respectively D-term and F-term variations. It is known that the first ones do not affect the vacuum geometry. An important consequence is that the computations for a $\mathcal{N}=(2,2)$ supersymmetric theory in two dimensions can be done in the corresponding $\mathcal{N}=4$ model in one dimension and the other way around. The argument is that we can choose a Lorentz-breaking Kahler potential which suppresses all the non-zero modes arising in the dimensional reduction to supersymmetric quantum mechanics \cite{rif4,rif14}. Thus, the indipendence of $tt^{*}$ geometry from the Kahler potential implies also indipendence from the dimensions.\\
The F-term deformation space is a complex manifold with holomorphic coordinates the complex couplings $t^{i}$ from which the superpotential depends. The tangent space is instead parametrized by chiral fields $ \Phi_{i}$ describing the F-term variations

\begin{equation*}
\int d^{2}\theta d^{2}z \ \delta W + c.c. = \int d^{2}\theta d^{2}z \ \delta t^{i} \Phi_{i}  + c.c. ,
\end{equation*}

where $\Phi_{i} = \partial_{t^{i}} W$. These operators and the anti-chiral ones form a commutative ring:

\begin{equation*}
\Phi_{i}\Phi_{j}= C_{ij}^{k} \Phi_{k}, \hspace{2cm}  \overline{\Phi}_{i} \ \overline{\Phi}_{j}= \overline{C}_{ij}^{k}\overline{\Phi}_{k},
\end{equation*}

where the structure constants $C_{ij}^{k}$ and $\overline{C}_{ij}^{k}$ depend holomorphically on $t^{i}$ and $\bar{t}^{i}$ respectively.\\ Another type of deformation that we can consider is the RG flow of the theory in the space of couplings. Even though the superpotential is protected by non-renormalization theorems, the F-term picks up a factor due to the rescaling of the superspace coordinates \cite{rif4}. From $z\rightarrow \mu z$ and $\theta \rightarrow \mu^{-1/2}\theta$ we get 

\begin{equation*}
\int d^{2}\theta d^{2}z \ W \longrightarrow \mu \int  d^{2}\theta d^{2}z \ W.
\end{equation*}

Variations in the overall factor $\mu$ generate a flow in the moduli space which has UV limit for $\mu \rightarrow 0$ and IR limit for $\mu \rightarrow \infty$.\\ The geometry of vacua is naturally studied in the topological twisted version of the theory \cite{rif15}. This is identical to the untwisted one on the flat space and contains as only physical BRST observables the operators in the chiral ring $\mathcal{R}=\frac{\mathbb{C}\left[ X_{i}\right] }{\partial_{j} W}$, where the fields $X_{i}$ are the fundamental degrees of freedom. In particular, there is a one to one correspondence between chiral operators $\Phi_{i}$ and ground states in the vacuum space. Denoting with $\ket{0}$ the state corresponding to the identity, we can define a frame of holomorphic sections on the vacuum bundle as

\begin{equation*}
\ket{i}= \Phi_{i} \ket{0}. 
\end{equation*}

In this basis of vacua the chiral operators are represented by the ring coefficients $C_{ij}^{k} $

\begin{equation*}
\Phi_{i}\ket{j}= C_{ij}^{k} \ket{k}, \hspace{2cm} (C_{i})_{j}^{k}= C_{ij}^{k}.
\end{equation*}

Similarly, the anti-topological theory provide a distinct basis corresponding to the anti-chiral fields $\ket{\overline{i}}$. Holomorphic and anti-holomorphic sections span the same vector space and are related by a matrix $M$:

\begin{equation*}
\ket{\overline{i}}= M_{\overline{i}}^{j} \ket{j},
\end{equation*}

which is also called real structure. \\ Now that the vacuum bundle is provided with an holomorphic structure, we can define two important quantities:

\begin{equation*}
\eta_{ij}= \braket{i}{j}, \hspace{2cm} g_{i\bar{j}}= \braket{\overline{j}}{i}.
\end{equation*}

$\eta$ is a symmetric pairing between chiral oparators and depends only on the holomorphic parameters. This is a purely topological object which can be easily computed with the techniques of the topological theory \cite{rif16}, or by using dimensional reduction to supersymmetric quantum mechanics \cite{rif17}.  One finds the formula

\begin{equation}
\eta_{ij}= \langle \Phi_{i} \Phi_{j} \rangle_{\mathrm{top.}} = \eta_{ij}= \braket{i}{j} = \mathrm{Res}_{W}\left[ \Phi_{i} \Phi_{j} \right] ,
\end{equation}

where $ \mathrm{Res}_{W}[\cdot]$ is the Groothendieck residue symbol defined by 

\begin{equation}
\mathrm{Res}_{W}[\Phi]= \frac{1}{(2\pi i)^{n}} \int_{\Gamma} \frac{\Phi(X) dX^{1} \wedge...\wedge dX^{n}}{\partial_{1}W\partial_{2}W...\partial_{n}W}.
\end{equation}

Instead, $g$ depends on both $t$ and $\bar{t}$ and represents an hermitian metric on the Hilbert space. Conversely to $\eta$, its computation requires much more efforts. The $3$ matrices $M,\eta$ and $g$ are related by 

\begin{equation*}
M=\eta^{-1}g.
\end{equation*}

Moreover, since $M$ plays the role of CPT operator of the theory, it must be satisfied the condition $MM^{*}=I$, which implies the `reality constraint'

\begin{equation}\label{realityconstraint}   
(\eta^{-1}g)(\eta^{-1}g)^{*}=I.
\end{equation}

The object studied by the $tt^{*}$ geometry is the Berry's connection of the vacuum bundle. This describes how the ground states transform under variations of the couplings and has the usual definition

\begin{equation*}
A_{ij\bar{k}}= \bra{\overline{k}}\partial_{i}\ket{j}, \hspace{1cm} A_{ij}^{k}= A_{ij\bar{l}}g^{k\bar{l}}, \hspace{1cm} A_{i \bar{j}}^{\bar{k}}= (A_{\bar{i}j}^{k})^{*},
\end{equation*}

with covariant derivatives

\begin{equation*}
D_{i}= \partial_{i}-A_{i}, \hspace{2cm} \overline{D}_{i}= \overline{\partial}_{i}-\overline{A}_{i}.
\end{equation*}

One can show that $A$ is metric with respect to $g$, i.e. $D_{i}g=\overline{D}_{i}g=0$, and is compatible with the holomorphic structure of the bundle. Thus, the Berry's connection is the Chern's connection of the vacuum bundle and in the holomorphic basis we have 

\begin{equation*}
A_{ij}^{k}=-g_{j\bar{l}}(\partial_{i}g^{-1})^{\bar{l}k}, \hspace{1cm} (A_{\bar{i}})_{j}^{k}=0.
\end{equation*}

In this frame the $tt^{*}$ geometry prescribes a differential equation for the ground state metric $g$ \cite{rif4}

\begin{equation}
\overline{\partial}_{i}(g\partial_{j}g^{-1})=\left[   C_{j}  , g (C_{i})^{\dagger}g^{-1} \right].
\end{equation}

So, if we want to study how the vacua transform under deformations of the theory, we have to solve this equation with the constraint \ref{realityconstraint}.\\ If the theory is massive, i.e. all the classical vacua are isolated, it is possible to identify an operator in $\mathcal{R}$ with the set of its values at the critical points. In particular, we can introduce a basis in which the chiral ring algebra diagonalizes completely, usually called `point basis '. This can be normalized such that 

\begin{equation*}
\Phi_{i}\ket{j}=\delta_{i}^{j} \ket{j}, \hspace{2cm} \Phi_{i}(\lbrace x\rbrace_{j})= \delta_{i}^{j},
\end{equation*}

where $\lbrace x\rbrace_{j}, j=1,...,\# \mathrm{ classical \ vacua}$ denote the classical vacuum configurations of the superpotential. In other words, for a massive theory $\mathcal{R}$ is $C^{\# \mathrm{ classical \ vacua}}$ as complex algebra. Also the matrices of the F-term chiral deformations diagonalize:

\begin{equation}
(C_{i})_{j}^{k}= \partial_{i}W(\left\lbrace x\right\rbrace_{j})\delta_{j}^{k}. 
\end{equation}

Moreover, in the massive case the formula of the Groothendieck residue becomes:

\begin{equation}\label{residue}   
\mathrm{Res}_{W}\left[ \Phi \right] = \sum_{\left\lbrace x\right\rbrace_{j}} \Phi(X) \mathfrak{H}^{-1},
\end{equation}

where $\mathfrak{H}=\mathrm{det}\partial_{i}\partial_{j}W$ is the Hessian of $W$.\\
Another prerogative of massive theories is that they posses a natural system of coordinates in the coupling constant space, i.e. the set of critical values of the superpotential $w_{i}=W(\lbrace x\rbrace_{i})$ \cite{rif5}. These are usually called canonical coordinates. As well as being convenient from a practical point of view, these parameters have a clear physical meaning, since the soliton masses of the theory are expressed as differences of critical values. Therefore, they represent an appropriate choice to investigate both the physical and mathematical properties of the ground state metric. In these coordinates the $tt^{*}$ equations become universal and different physical systems are distinguished by the boundary conditions imposed on the solutions.

\subsubsection{$tt^{*}$ Equations}\label{trunc}

Our next aim is to derive the $tt^{*}$ equations for this class of models. We first introduce the point basis $\Phi_{k;m,n,n_{0},...,n_{N-2}}$ defined by

\begin{equation}\label{rep}
\Phi_{k;m,n,n_{0},...,n_{N-2}} ( p_{k^{\prime};m^{\prime},n^{\prime},n_{0}^{\prime},...,n_{N-2}^{\prime}};\tau) =\delta_{k,k^{\prime}}\delta_{m,m^{\prime}} \delta_{n,n^{\prime}} \prod_{r=0}^{N-2} \delta_{n_{r},n_{r}^{\prime}}.
\end{equation}

The isomorphism between $\mathcal{R}$ and $\mathbb{C}^{\# \mathrm{ classical \ vacua}}$ is realized by the identification

\begin{equation*}
\Phi_{k;m,n,n_{0},...,n_{N-2}}\longrightarrow \ket{p_{k;m,n,n_{0},...,n_{N-2}}}= A^{m} B^{ n} \prod_{r=0}^{N-2} L_{r}^{  n_{r}} \sigma^{ k}\ket{p_{0}},
\end{equation*}

where the operatorial action of the symmetry algebra is naturally transported on the Hilbert space. The vacuum space of the theory on $\mathcal{S}$ decomposes in a direct sum of irreducible representations of the homology group:

\begin{equation*}
\mathcal{V}_{\mathcal{S}} \simeq \mathrm{L}^{2}\left(  \mathrm{Hom} \left(   \mathbb{Z}^{N+1}, U(1) \right) \right) \otimes \mathbb{C}^{N},
\end{equation*}

provided that $N+1$ is the rank of $H_{1}(\mathcal{S},\mathbb{Z})$ and $N$ the number of vacua on $\mathcal{S}$. A point basis of `theta-vacua' for $\mathcal{V}_{\mathcal{S}}$ is 

\begin{equation}\label{point}
\begin{split}
 \ket{k;\alpha, \beta, \lambda_{0},...,\lambda_{N-2}}=& \ e^{\frac{-i \alpha k}{N}} \sum_{m,n,n_{r} \in \mathbb{Z}} e^{-i( m\alpha + n\beta+\sum_{r=0}^{N-2}n_{r}\lambda_{r} )} \\ \\ & A^{m} B^{n} \prod_{r=0}^{N-2} L_{r}^{ n_{r}} \sigma^{ k}\ket{p_{0}},
\end{split}
\end{equation}

where the angles $\alpha, \beta, \lambda_{0},...,\lambda_{N-2} \in [0,2\pi]$ label representations of $H_{1}(\mathcal{S},\mathbb{Z})$. Since the homology is abelian, the angles are defined simultaneously in this common basis of eigenstates:

\begin{equation*}
\begin{split}
& A\ket{k;\alpha, \beta, \lambda_{0},...,\lambda_{N-2}}=e^{i\alpha} \ket{k;\alpha, \beta, \lambda_{0},...,\lambda_{N-2}}, \\ \\ & B\ket{k;\alpha, \beta, \lambda_{0},...,\lambda_{N-2}}=e^{i\beta} \ket{k;\alpha, \beta, \lambda_{0},...,\lambda_{N-2}}, 
\\ \\ & L_{r}\ket{k;\alpha, \beta, \lambda_{0},...,\lambda_{N-2}}=e^{i\lambda_{r}} \ket{k;\alpha, \beta, \lambda_{0},...,\lambda_{N-2}}.
\end{split}
\end{equation*}

Using the symmetries $A,B,L_{r}$, one finds that in the point basis the ground state metric diagonalizes with respect to the angles:

\begin{equation*}
\begin{split}
& \braket{\overline{j;\alpha^{\prime}, \beta^{\prime}, \lambda_{0}^{\prime},...,\lambda_{N-2}^{\prime}}}{k;\alpha, \beta, \lambda_{0},...,\lambda_{N-2}}= \\ \\ & \delta(\alpha-\alpha^{\prime})\delta(\beta-\beta^{\prime}) \prod_{r=0}^{N-2} \delta( \lambda_{r}-\lambda_{r}^{\prime}) \  g_{k,\bar{j}}( \alpha, \beta, \lambda_{0},...,\lambda_{N-2}), 
\end{split} 
\end{equation*}

where the matrix coefficients $g_{k,\bar{j}}( \alpha, \beta, \lambda_{0},...,\lambda_{N-2})$ have the Fourier expansion:

\begin{equation*}
\begin{split}
& g_{k,\bar{j}}( \alpha, \beta, \lambda_{0},...,\lambda_{N-2}) =  e^{\frac{i}{N}\alpha (j-k)}\sum_{r,s,t_{q} \in \mathbb{Z}} e^{i( r\alpha + s\beta+\sum_{q=0}^{N-2}t_{q}\lambda_{q} )} g_{k,\bar{j}}(r,s,t_{0},...,t_{N-2}), \\ \\ 
& g_{k,\bar{j}}(r,s,t_{0},...,t_{N-2})= \braket{\overline{p_{j;m,n,n_{0},...,n_{N-2}}}}{p_{k;m^{\prime},n^{\prime},n_{0}^{\prime},...,n_{N-2}^{\prime}}} \vert_{m-m^{\prime}=r, n-n^{\prime}=s, n_{q}-n_{q}^{\prime}=t_{q}} .
\end{split}
\end{equation*}

The metric cannot be diagonalized also as $N \times N$ matrix, since $\sigma$ do not commute with $B$ and $L_{r}$. In general we are interested in solutions to the $tt^{*}$ equations which are compatible with the abelianity of the model projected on the punctured plane. Hovewer, we cannot just set the loop angles to be $0$, since these are differential variables which enter in the equations. So, first we have to check that the ansatz of a solution with constant vanishing $\lambda_{r}$ is consistent with all the equations. From the equality 

\begin{equation*}
\frac{2 \pi i}{e^{\frac{2\pi i l}{N}}-1}= \frac{2\pi i}{N} \sum_{k=0}^{N-1} k e^{\frac{2\pi i l k}{N}}= \frac{2\pi i}{N} \sum_{k=0}^{N-2} (k-N+1) e^{\frac{2\pi i l k}{N}},
\end{equation*}

we learn that the combination $\tilde{B}= B^{N}\prod\limits_{k=0}^{N-2} L_{k}^{N-k-1}$ leaves invariant the superpotential. Moreover, one can see that it commutes with $\sigma$

\begin{equation*}
\begin{split}
&\sigma \tilde{B}= \sigma B^{N}\prod_{k=0}^{N-2} L_{k}^{N-k-1}= B^{N}L_{0}^{N}\prod_{k=0}^{N-3} L_{k+1}^{N-k-1} \prod_{k=0}^{N-2} L_{k}^{-1} \sigma = \\ & B L_{0}^{N-1} \prod_{k=0}^{N-3} L_{k+1}^{N-(k+1)-1}\sigma = B^{N}\prod\limits_{k=0}^{N-2} L_{k}^{N-k-1}\sigma = \tilde{B}\sigma
\end{split}
\end{equation*}

where we have used the commutation relations in \ref{identity}. Thus, if we choose $A,\tilde{B},L_{r}$ as generators of the homology, we can find a basis in which the eigenvalues of $\sigma$ are simultaneously defined with the two angles $\alpha,\tilde{\beta}= N\beta + \sum_{k=0}^{N-2}(N-k-1)\lambda_{k}$ corresponding to $A,\tilde{B}$.\\ The couplings of which we can consider variations in these models are the lattice parameter $\tau$ and the overall scale $\mu$. The corresponding chiral ring coefficients in the point basis of the covering model read

\begin{equation}
\begin{split}
\left( C_{\tau}\right)_{ k;m,n,n_{0},...,n_{N-2}}^{k^{\prime};m^{\prime},n^{\prime},n_{0}^{\prime},...,n_{N-2}^{\prime}}= & \
 \mu \partial_{\tau} W^{(N,l)}\Phi_{k^{\prime};m^{\prime},n^{\prime},n_{0}^{\prime},...,n_{N-2}^{\prime}} (p_{k;m,n,n_{0},...,n_{N-2}};\tau) \\  = &  \ \delta_{k,k^{\prime}}\delta_{m,m^{\prime}} \delta_{n,n^{\prime}} \prod_{r=0}^{N-2} \delta_{n_{r},n_{r}^{\prime}} \ \mu \partial_{\tau} W^{(N,l)}(p_{0};\tau)e^{\frac{2\pi i l k}{N}}, \\ \\ 
\left( C_{\mu}\right)_{ k;m,n,n_{0},...,n_{N-2}}^{k^{\prime};m^{\prime},n^{\prime},n_{0}^{\prime},...,n_{N-2}^{\prime}}= & 
 \ W^{(N,l)}\Phi_{k^{\prime};m^{\prime},n^{\prime},n_{0}^{\prime},...,n_{N-2}^{\prime}}   (p_{k;m,n,n_{0},...,n_{N-2}};\tau) \\ = & \ \delta_{k,k^{\prime}}\delta_{m,m^{\prime}} \delta_{n,n^{\prime}} \prod_{r=0}^{N-2} \delta_{n_{r},n_{r}^{\prime}} \\ &  \mathrm{x} \ \left( W^{(N,l)}(p_{0};\tau)e^{\frac{2\pi i l k}{N}} +\sum_{r=0}^{N-2} n_{r} \ 2\pi i \ e^{\frac{2\pi i l r}{N}} \right), 
\end{split}
\end{equation}

where we have replaced $B$ with $\tilde{B}$ in the computation of the critical values in \ref{critical}.\\ The action of $C_{\tau}$ and $C_{\mu}$ is naturally projected on the theta-sectors of $\mathcal{V}_{\mathcal{S}}$:

\begin{equation}
\begin{split}
 C_{\tau}^{\mathrm{point \ basis}}(\alpha,\tilde{\beta} , \lambda_{0},...,\lambda_{N-2})_{j,k}=& \ \mu \partial_{\tau} W^{(N,l)}(p_{0};\tau) e^{\frac{2\pi i l k}{N}} \delta_{j,k}, \\ \\ 
 C_{\mu}^{\mathrm{point \ basis}}(\alpha, \tilde{\beta}, \lambda_{0},...,\lambda_{N-2})_{j,k}=& \left( W^{(N,l)}(p_{0};\tau)e^{\frac{2\pi i l k}{N}}- 2\pi \sum_{r=0}^{N-2}e^{\frac{2\pi i l r}{N}} \frac{\partial}{\partial \lambda_{r}} \right) \delta_{j,k}.
\end{split}
\end{equation}

The ground state metric satisfy the following set of equations 

\begin{equation}\label{set}
\begin{split}
& \partial_{\bar{\tau}}( g \partial_{\tau} g^{-1})= \left[C_{\tau}  ,g C_{\tau}^{\dagger}g^{-1}\right], \\ \\ 
& \partial_{\bar{\mu}}( g \partial_{\mu} g^{-1})= \left[C_{\mu}  ,g C_{\mu}^{\dagger}g^{-1}\right], \\ \\ 
& \partial_{\bar{\mu}}( g \partial_{\tau} g^{-1})= \left[C_{\tau}  ,g C_{\mu}^{\dagger}g^{-1}\right],
\end{split}
\end{equation}

as well as the complex conjugates. If we demand that $\lambda_{0}=...=\lambda_{N-2}=0$ and $\frac{\partial g}{\partial \lambda_{r}}=0$ for each $r$, the derivatives $\frac{\partial}{\partial \lambda_{r}}$ do not contribute to the $tt^{*}$ equations and the F-term deformations can be truncated at the non trivial part 

\begin{equation}
\begin{split}
 C_{\tau}^{\mathrm{point \ basis}}(\alpha, \tilde{\beta})_{j,k}=& \mu \partial_{\tau} W^{(N,l)}(p_{0};\tau) e^{\frac{2\pi i l k}{N}} \delta_{j,k}, \\ \\ 
 C_{\mu}^{\mathrm{point \ basis}}(\alpha, \tilde{\beta})_{j,k}=&  W^{(N,l)}(p_{0};\tau)e^{\frac{2\pi i l k}{N}} \delta_{j,k},
\end{split}
\end{equation}

with $\tilde{\beta}=N\beta$. It is clear that with this ansatz the equations given above can be written universally as a unique equation with respect to the critical value $w=W^{(N,l)}(p_{0};\tau)$:

\begin{equation}\label{canonical}   
\partial_{\bar{w}}( g \partial_{w} g^{-1})= \left[C_{w}  , g C_{w}^{\dagger}g^{-1}\right],
\end{equation}

where 

\begin{equation}
C_{w}^{\mathrm{point \ basis}}(\alpha, \tilde{\beta})_{j,k}= e^{\frac{2\pi i l k}{N}} \delta_{j,k}.
\end{equation}

The equations in the parameters $\tau$ and $\mu$ can be obtained from this one by specifying the variation of $w$ and $\bar{w}$. Therefore, if the equation in the canonical variable is solved with the boundary conditions given by the cusps, the solution automatically satisfies all the equations in \ref{set}. By consistency with the critical limit $\mu \rightarrow 0$, one of the UV cusp must be located in the origin of the $W$-plane. In conclusion, it is convenient to pull-back the equation on the spectral cover of the model where one can use $w= W^{(N,l)}(p_{0};\tau)$ as unique complex coordinate to parametrize models and vacua.\\ Now that we have consistently recovered the abelian representation of the symmetry group, we can construct a common basis of eigenstates for $\sigma,A,B$ in which the ground state metric diagonalizes completely. Given the action of $\sigma$ on the point basis 

\begin{equation*}
\sigma\ket{k;\alpha, \beta}=e^{\frac{i \alpha }{N}} \ket{k+1;\alpha, \beta},
\end{equation*}

we can define a set of $\sigma$-eigenstates as 

\begin{equation}\label{eigen}
\begin{split}
& \ket{j;\alpha, \beta}= \sum_{k=0}^{N-1}  e^{- \frac{2 \pi i l k j}{N}}  \ket{k;\alpha, \beta}, 
\\ \\ & \sigma \ket{j;\alpha, \beta}= e^{\frac{ i}{N} (\alpha + 2\pi l j)} \ket{j;\alpha, \beta}.
\end{split}
\end{equation}

With this normalization we have the periodicity 

\begin{equation}
\ket{k+N;\alpha, \beta}=\ket{k;\alpha, \beta}
\end{equation}

and we note that the given definitions are consistent with the relation $\sigma^{N}=A$. \\ In the $\sigma$-basis the metric becomes

\begin{equation}
 g_{k,\bar{j}}( \alpha, \beta) = \delta_{k,j} \ e^{\varphi_{k}( \alpha, \beta)},
\end{equation}

where the $\varphi_{k}( \alpha, \beta), k=0,..,N-1,$ are real functions of the angles. It is straightforward to derive the $tt^{*}$ equations satisfied by these functions. Using

\begin{equation*}
C_{w}^{\sigma-\mathrm{basis}}(\alpha, \beta)= 
\begin{pmatrix}  
0 & 1 & 0 & . . . & 0 & 0 \\ 
0 & 0 & 1 &. . . & 0 & 0 \\
. & . &  . &     & . & . \\ 
. & . &  . &     & . & . \\ 
 0 & 0 & 0 & . . . & 0 & 1 \\ 
 1 & 0 & 0 & . . . & 0 & 0 \\
\end{pmatrix},
\end{equation*}

one finds the well known $\hat{A}_{N-1}$ Toda equations:

\begin{equation}\label{toda} 
\begin{split}
& \partial_{\bar{w}}\partial_{w} \varphi_{0} + e^{\varphi_{1}-\varphi_{0}}-e^{\varphi_{0}-\varphi_{N-1}}=0 ,\\ \\
& \partial_{\bar{w}}\partial_{w} \varphi_{i} + e^{\varphi_{i+1}-\varphi_{i}}-e^{\varphi_{i}-\varphi_{i-1}}=0, \hspace{2cm} i=1,...,N-2 \\ \\
& \partial_{\bar{w}}\partial_{w} \varphi_{N-1} + e^{\varphi_{0}-\varphi_{N-1}}-e^{\varphi_{N-1}-\varphi_{N-2}}=0,
\end{split}
\end{equation}

where the dependence on the angles and $w$ is understood. These equations appear tipically in the models with a $\mathbb{Z}_{N}$ symmetry group acting transitively on the vacua. \\ We can also provide an expression for the symmetric pairing of the topological theory. The operators in the chiral ring corresponding to the basis in \ref{point} with $\lambda_{r}=0$ and \ref{eigen} are respectively

\begin{equation*}
\begin{split}
& \chi_{k}(\alpha, \beta)= e^{\frac{-i \alpha k}{N}} \sum_{m,n \in \mathbb{Z}} e^{-i( m\alpha + n\beta)} \Phi_{k;m,n}, \\ \\ 
& \Psi_{j}(\alpha, \beta)= \sum_{k=0}^{N-1}  e^{- \frac{2 \pi i l k j}{N}}  \chi_{k}(\alpha, \beta).
\end{split}
\end{equation*}

The topological metric can be computed with the formula \ref{residue} for one chiral superfield. In the point basis the expression is

\begin{equation}
\begin{split}
\mathrm{Res}_{W} \left[ \chi_{k}(\alpha, \beta), \chi_{m}(\alpha^{\prime}, \beta^{\prime}) \right] = \delta(\alpha-\alpha^{\prime})\delta(\beta-\beta^{\prime}) \eta_{k,m}^{\mathrm{point \ basis}}(\alpha, \beta)  
\end{split}   
\end{equation}

where 

\begin{equation}
\eta_{k,m}^{\mathrm{point \ basis}}(\alpha, \beta) = \left( \partial^{2} _{z} W^{(N,l)}(0;\tau)\right) ^{-1} e^{-\frac{ i k}{N} (2\alpha + 2\pi l )}\delta_{k,m} .
\end{equation}

In the basis of $\sigma$-eigenstates one gets

\begin{equation}
\begin{split}
& \mathrm{Res}_{W} \left[ \Psi_{n}(\alpha, \beta), \Psi_{m}(\alpha^{\prime}, \beta^{\prime}) \right] =  \delta(\alpha-\alpha^{\prime})\delta(\beta-\beta^{\prime})\ \eta_{n,m}^{\sigma-\mathrm{basis}}(\alpha, \beta), 
\end{split}
\end{equation}

where

\begin{equation}
\eta_{n,m}^{\sigma-\mathrm{basis}}(\alpha, \beta)=\left( \partial^{2} _{z} W^{(N,l)}(0;\tau)\right) ^{-1}  \sum_{k=0}^{ N-1} e^{-\frac{ i k}{N} (2\pi l (m+n+1) + 2\alpha )} .
\end{equation}

We can choose a basis and use one of these expressions to impose the reality constraint \ref{realityconstraint},
where the complex conjugates $g^{*}, \eta^{*}$ in terms of the matrices $g(\alpha,\beta)$, $\eta(\alpha,\beta)$ are respectively 

\begin{equation*}
\begin{split}
& g^{*}(\alpha, \beta)=\left[  g(-\alpha,- \beta) \right]  ^{*}, \\ & \eta^{*}(\alpha, \beta)=\left[  \eta(-\alpha, -\beta) \right]^{*}.
\end{split}
\end{equation*}

This condition gives $g(-\alpha, -\beta)$ in terms of $g(\alpha, \beta)$, but an explicit computation is hard in general. In the case of a trivial representation of the homology, the reality constraint simplifies, becoming the same of the $A_{N}$ models \cite{rif4}.

\subsubsection{Peculiarities of the N=2 Level}

We briefly discuss some peculiar aspects of the class of models of level $2$. The derivative of the superpotential is 

\begin{equation}
\partial_{z} W(z;\tau)= \zeta ( z-\pi \tau;\tau)-\zeta(z-\pi \tau -\pi ;\tau) -\eta_{1},
\end{equation}

where  $  z \in \mathcal{S}= \mathbb{C} \setminus \left\lbrace \pi \tau, \pi \tau +\pi +\Lambda_{\tau}; \  \Lambda_{\tau}= 2\pi\mathbb{Z} \oplus 2\pi \tau \mathbb{Z}\right\rbrace $.  The simple poles and simple zeroes are located respectively in $\pi \tau, \pi \tau +\pi +\Lambda_{\tau}$ and $0, \pi+\Lambda_{\tau}$. In this case the Galois group acts trivially and we have only one co-level.\\ In addition to the generators $\sigma, A, B,L$ already discussed, the symmetry group of the models of level $2$ contains also the parity transformation 

\begin{equation*}
\iota : z\longrightarrow -z; \hspace{2cm} \partial_{z} W(-z;\tau)=- \partial_{z} W(z;\tau),
\end{equation*}

which follows from \ref{colevel} for $N=2$ and $l=1$. \\ This operator satisfies the following commutation relations with the generators of the abelian group:

\begin{equation*}
\begin{split}
& \iota A=  A^{-1} \iota, \\ 
& \iota B=  B^{-1} \iota, \\ 
& \iota L=  L \iota, \\ 
& \iota \sigma=  \sigma^{-1} \iota .
\end{split}
\end{equation*}

The action of $ \iota$, as well as its algebraic relations, can be extended to the abelian universal cover. There are inequivalent choices which differ by compositions with the loop generator $L$. Indipendently from the definition, the set of curves with $p(1)=0$ is left invariant by the action of $\iota$.\\ The presence of an extra symmetry, as well as the possibility of working with $2 \times 2$ matrices, guarantees some semplifications in the derivation of the $tt^{*}$ equations. A point basis of theta-vacua for this family of theories is
 
\begin{equation*}
\begin{split}
& \ket{0;\alpha, \beta, \lambda}= \sum_{m,n,j \in \mathbb{Z}} e^{-i( m\alpha + n\beta+j\lambda)} A^{m} B^{  n} L^{j} \ket{p_{0}}, 
\\ &
\ket{1;\alpha, \beta, \lambda}= \sum_{m,n,j \in \mathbb{Z}} e^{-i( m\alpha + n\beta+j\lambda)} A^{m} B^{  n} L^{j} \sigma\ket{p_{0}}.
\end{split}
\end{equation*}

Setting $\lambda=0$, the operators $\sigma$ and $\iota$ act on these states as:

\begin{equation*}
\begin{split}
& \sigma \ket{0;\alpha, \beta}= \ket{1;\alpha, \beta}, \hspace{2cm} \sigma \ket{1;\alpha, \beta} = e^{i \alpha} \ket{1;\alpha, \beta}, \\ 
& \iota \ket{0;\alpha, \beta}= \ket{0;-\alpha, -\beta}, \hspace{1.5cm} \iota \ket{1;\alpha, \beta} = e^{-i \alpha} \ket{1;-\alpha, -\beta}.
\end{split}
\end{equation*}

The ground state metric in this basis is represented by the $2$ x $2$ hermitian matrix

\begin{equation*}
g(\alpha, \beta)= \begin{pmatrix}
g_{0\bar{0}}(\alpha, \beta) & g_{0\bar{1}}(\alpha, \beta) \\ g_{1\bar{0}}(\alpha, \beta) & g_{1\bar{1}}(\alpha, \beta)
\end{pmatrix}.
\end{equation*}

The symmetries $\sigma$ and $\iota $ imply respectively

\begin{equation*}
g_{0\bar{0}}(\alpha, \beta)= g_{1\bar{1}}(\alpha, \beta), \hspace{2cm} 
g_{0\bar{1}}(\alpha, \beta)= e^{i\alpha} g_{1\bar{0}}(\alpha, \beta),
\end{equation*}

and

\begin{equation*}
g_{0\bar{0}}(\alpha, \beta)= g_{0\bar{0}}(-\alpha, -\beta), \hspace{2cm} 
g_{1\bar{0}}(\alpha, \beta)= e^{-i\alpha} g_{1\bar{0}}(-\alpha, -\beta).
\end{equation*}

Therefore the metric can be written as 

\begin{equation*}
g(\alpha, \beta)= \begin{pmatrix}
A(\alpha, \beta) & e^{\frac{i\alpha}{2}}B(\alpha, \beta) \\ e^{\frac{-i\alpha}{2}}B(\alpha, \beta) & A(\alpha, \beta)
\end{pmatrix},
\end{equation*}

where $A(-\alpha, -\beta)=A(\alpha, \beta)$ and $B(-\alpha, -\beta)=B(\alpha, \beta)$.  The transpose and complex conjugate of $g$ read  

\begin{equation*}
\begin{split}
& g^{\mathrm{T}}(\alpha, \beta)=\left[  g(-\alpha,- \beta) \right]  ^{\mathrm{T}}, \\ & g^{*}(\alpha, \beta)=\left[  g(-\alpha, -\beta) \right]^{*},
\end{split}
\end{equation*}

and therefore 

\begin{equation*}
g^{\mathrm{\dagger}}(\alpha, \beta)= \left[  g(-\alpha,- \beta) \right]  ^{\mathrm{\dagger}}.
\end{equation*}

The hermiticity of $g$ implies that $A(\alpha, \beta)$ and $B(\alpha, \beta)$ are real functions, while the positivity requires that $A(\alpha, \beta) > 0$. The matrix $C_{w}$ and the topological metric in this basis are 

\begin{equation}
C_{w}(\alpha, \beta)= \begin{pmatrix}
1 & 0 \\ 0 & -1
\end{pmatrix}, \hspace{2cm }
\eta(\alpha, \beta)= \left( \partial^{2} _{z} W(0;\tau)\right) ^{-1} \begin{pmatrix}
1 & 0 \\ 0 & -1
\end{pmatrix}.
\end{equation}

We see that, by the parity properties of $A(\alpha, \beta)$ and $B(\alpha, \beta)$ implied by $ \iota$,  the reality constraint \ref{realityconstraint} reduces to 

\begin{equation}
\vert \partial^{2} _{z} W(0;\tau) \vert ^{2} ( A(\alpha, \beta)^{2}-B(\alpha, \beta)^{2} )=1.
\end{equation}

As a conquence, the metric can be parametrized in term of a single function of the angles

\begin{equation}
\begin{split}
& A(\alpha, \beta)= \frac{1}{\vert \partial^{2} _{z} W(0;\tau) \vert} \cosh \left[ L(\alpha, \beta)\right], \\ 
& B(\alpha, \beta)= \frac{1}{\vert \partial^{2} _{z} W(0;\tau) \vert} \sinh \left[ L(\alpha, \beta)\right] .
\end{split}
\end{equation}

Finally, by plugging the given expressions of $g(\alpha, \beta)$ and $C_{w}$ in \ref{canonical} we find the Sinh-Gordon equation

\begin{equation}
\partial_{\bar{w}}\partial_{w}  L(\alpha, \beta) = 2 \sinh \left[ 2 L(\alpha, \beta)\right].
\end{equation}

\section{Modular Properties of the Models}

\subsection{Modular Transformations of $ \partial_{z} W^{(N,l)}(z;\tau)$}

The underlying structure of this class of models is the theory of the modular curves. In particular, as discussed in \ref{modular}, the space of models and its spectral cover are described respectively by the modular curves for $\Gamma_{1}(N)$ and $\Gamma(N)$. Now that we have provided an explicit description of these systems, we want to investigate the modular properties of the superpotential, as well as the ground state metric, with respect to the relevant congruence subgroups of $SL(2,\mathbb{Z})$. Let us first consider $\partial_{z} W^{(N,l)}(z;\tau)$. Given a modular transformation $\gamma = \begin{pmatrix} a & b \\ c & d \end{pmatrix} \in \Gamma_{1}(N)$, the derivative of the superpotential transforms as 

\begin{equation}
\begin{split}
& \partial_{z} W^{(N,l)}\left( \frac{z}{c\tau + d};\frac{a\tau +b}{c\tau + d}\right) = \\ \\ &
 (c\tau+d) \sum_{k=0}^{N-1} e^{\frac{2\pi i l k}{N}} \biggl[ \zeta \left( z- \frac{2\pi}{N}((a\tau+b)l + k (c\tau + d)) ;\tau \right) + 2\frac{d \eta_{1}+c \eta_{2}}{N} k \biggr] = \\ \\ 
&(c\tau+d) \sum_{k=0}^{N-1} e^{\frac{2\pi i l k}{N}}  \left[ \zeta \left( z-\frac{2\pi}{N}bl- \frac{2\pi}{N}(l\tau+k) ;\tau \right) +\frac{2\eta_{1}k}{N}  \right]= \\ \\ & (c\tau+d)  \partial_{z} W^{(N,l)}\left( z-\frac{2\pi}{N}bl ;\tau\right),
\end{split}
\end{equation}

where we have used the fact that $a,d=1$ mod $N$, $c=0$ mod $N$ and the modular properties of the Weierstrass function: 

\begin{equation}\label{weierstrass}
\begin{split}
& \zeta \left(\frac{z}{c\tau + d} ; \frac{a\tau +b}{c\tau + d} \right)= (c\tau+d)\zeta(z;\tau), \\ \\ & \eta_{1}\left( \frac{a\tau +b}{c\tau + d}\right) = (c\tau+d) (d \eta_{1}+c \eta_{2}). 
\end{split}
\end{equation}

We see that $\Gamma_{1}(N)$ preserves only one of the torsion point. The transformation $\frac{2\pi\tau}{N} \rightarrow  \frac{2\pi(a\tau+b)}{N}$ results in a shift of poles and vacua by $\frac{2\pi}{N}bl$. If we set $b=0$ mod $N$, i.e. $\gamma \in \Gamma(N)$, also the order of vacua is preserved modulo periodicity.\\ On the other hand, a matrix $\gamma = \begin{pmatrix} a & b \\ c & d \end{pmatrix} \in \Gamma_{0}(N)$  acts on $\partial_{z} W^{(N,l)}(z;\tau)$ as 

\begin{equation}\label{dw}
\begin{split}
& \partial_{z} W^{(N,l)}\left( \frac{z}{c\tau + d};\frac{a\tau +b}{c\tau + d}\right) = \\ \\  
& (c\tau+d) \sum_{k=0}^{N-1} e^{\frac{2\pi i l k}{N}} \left[ \zeta \left( z- \frac{2\pi}{N}((a\tau+b)l + k (c\tau + d)) ;\tau \right) +2\frac{d \eta_{1}+c \eta_{2}}{N} k \right]= \\ \\ 
& (c\tau+d) \sum_{k=0}^{N-1} e^{\frac{2\pi i al k}{N}} \left[ \zeta \left( z-\frac{2\pi}{N}bl- \frac{2\pi}{N}(al\tau+k) ;\tau \right) +\frac{2\eta_{1}k}{N}  \right]=
\\ \\ & (c\tau+d) \partial_{z} W^{(N,al)}\left( z-\frac{2\pi}{N}bl ;\tau\right),
\end{split}
\end{equation}

where in the third line we used the fact that $a=d^{-1}$ mod $N$. As discussed in section \ref{modular}, $ \Gamma_{0}(N)$ changes the co-level with the map $l \rightarrow al$, acting as the Galois group of the real cyclotomic extension.

\subsection{Modular Transformations of the Physical Mass}

It turns out that the superpotential is not invariant under a transformation of $\Gamma(N)$, but it shifts by a constant. On one side, it is not wrong to say that the model is left invariant, since a constant is irrilevant in determining the physics of a system. But, at the same time,  the `physical mass' $W^{(N,l)}(p_{0};\tau)$ is the parameter that we have used to write the $tt^{*}$ equations. This means that $\Gamma(N)$ changes the coordinate of the model on the spectral cover, with a consequent transformation of the ground state metric. An analogous constant shift is induced on the superpotential also by $\Gamma_{1}(N)$ and $\Gamma_{0}(N)$, in addition to the effects discussed in the previous paragraph. Before giving an interpretation of such phenomena and solving the apparent contraddiction, we want to compute these constants.\\ The fact that $W^{(N,l)}(p_{0};\tau)$ transforms non trivially under $\Gamma(N)$ is connected to the geometry of the modular curves. These spaces are not simply connected and a critical value can be defined only on the universal cover, i.e. the upper half plane, where $\Gamma(N)$ plays the role of deck group. This is also strictly related to the fact that the superpotential is defined on the universal cover of the target space, where we have more than $N$ vacua. On the other hand, variations of $W^{(N,l)}(p_{0};\tau)$ with respect to some coordinate on the modular curve must be modular functions of $\Gamma(N)$. Let us give some examples. Among the curves $X(N)$, the cases with $N\leq 5$ are the only ones with genus $0$ and therefore the easiest to describe \cite{rif13}. Genus $0$ means in particular that the modular function field has a unique generator, also called Hauptmodul. This function is unique up to Mobius transformations and modular with respect to $\Gamma(N)$. Moreover, it is the projective coordinate which realizes the isomorphism between $X(N)$ and the Riemann sphere punctured with the cusps. Denoting with $x$ the coordinate on the sphere, we can define the one-form $dW(x)=C(x)dx$, where $C(x)=\partial_{x}W(p_{0};x)$ is the chiral ring coefficient which describes variations of the critical value with respect to the Hauptmodul. Except the case of $N=5$, for these curves we have only the co-level $1$ (and the inverse), so we suspend the notation $(N,l)$. From the theory of projective algebraic curves, $C(x)$ must be a rational function with coefficients in the cyclotomic field. Moreover, the poles must be located in the free IR cusps, where the physical mass is expected to diverge. Thus, $C(x)$ can be fixed up to a multiplicative constant by requiring the correct transformation properties under $\Gamma_{1}(N)$ and $\Gamma_{0}(N)$. In particular, from the transformations of $\partial_{z} W^{(N,l)}(z;\tau)$ under $\Gamma_{1}(N)$, we learn that  $T: \tau \rightarrow \tau +1$ acts on the one form as an automorphism of the sphere with the formula 

\begin{equation*}
T^{*}dW(x)= dW(p_{0};T^{*}x)=dW(\sigma^{-1}(p_{0});x)=e^{-\frac{2\pi i }{N}} dW(x),
\end{equation*}

where $l=1$ is understood.\\ Starting from $X(2)$, we have $3$ cusps which can be viewed as the $3$ vertices of the equatorial triangle of a double triangular pyramid inscribed in $\mathbb{P}^{1}$. A set of representatives for the cusps is given by 

\begin{equation*}
C_{\Gamma(2)}=\left\lbrace 0,1,\infty \right\rbrace 
\end{equation*}

where $\infty$, being a fixed point of $\tau \rightarrow \tau +1$, is a UV cusp and $0,1$ identify two decoupled vacua. Moreover, the quotient group $PSL(2,\mathbb{Z})/\Gamma(2)$ is isomorphic to $\mathbb{S}_{3}$, which is the symmetry group of a triangle. We can use $x(\tau)=1-\lambda(\tau)$ as coordinate on the sphere, where $\lambda(\tau): \mathbb{H}/\Gamma(2) \overset{\sim}{\longrightarrow} \mathbb{P}^{1}/\lbrace 0,1, \infty \rbrace$ is the modular lambda function defined by 

\begin{equation*}
\lambda(\tau)= \frac{\theta_{2}^{4}(0;\tau)}{\theta_{3}^{4}(0;\tau)}.
\end{equation*}

The $3$ cusps fall in the point $x(\infty)=1, x(0)=0,x(1)= \infty$ and $T$ acts on $x$ by the transformation of $PSL(2,\mathbb{C})$

\begin{equation*}
T^{*}x =\frac{1}{x}.
\end{equation*}

From what we have said $C(x)$ must have a simple pole in $x=0$ and $x=\infty$. Asking that $T^{*}dW(x)= -dW(x)$, we find 

\begin{equation}
dW(x)= \alpha \frac{dx}{x}
\end{equation}

where $\alpha$ is a complex constant. We see that there is a unique generator of $\Gamma(2)$ which acts non trivially on the critical value. This generates loops around the cusp in $0$ (as well as in $\infty$) and produces the constant shift of the superpotential.\\
In the case of $N=3$ the cusps are located at the $4$ vertices of a regular tetrahedron inscribed in the Riemann sphere. It is not a coincidence that $PSL(2,\mathbb{Z})/\Gamma(3)$ is isomorphic to the symmetry group of this solid figure, i.e. $\mathbb{A}_{4}$. The set of cusps is 

\begin{equation*}
C_{\Gamma(3)}=\left\lbrace 0,1/2,1,\infty\right\rbrace .
\end{equation*}

Also in this case $\infty$ is the only UV cusp, while the other rationals are in the same orbit of $T$ and identify free theories. The integers coprime with $3$ are $1$ and $2 =-1$ mod $3$ which describe the same theory. The generator of the function field is 

\begin{equation*}
J_{3}(\tau)= \frac{1}{i\sqrt{27}}q_{\tau}^{-\frac{1}{3}} \left( \prod_{n=1}^{\infty} \frac{1-q_{\tau}^{n/3}}{1-q_{\tau}^{3n}} \right)^{3} = \frac{1}{i\sqrt{27}} \left( \frac{\eta(\tau/3)}{\eta(3\tau)}\right)^{3} 
\end{equation*}

where $q_{\tau}=e^{2\pi i \tau}$ and 

\begin{equation}
\eta(\tau)= q_{\tau}^{1/24}  \prod _{n=1}^{\infty} (1-q_{\tau}^{n})
\end{equation}

is the Dedekind eta function. The generator of $\Gamma_{1}(3)/ \Gamma(3)$ acts on the Hauptmodul as 

\begin{equation*}
T^{*} J_{3}=\zeta_{3} + \zeta_{3}^{2}J_{3}
\end{equation*}

where we use the notation $ \zeta_{k}= e^{2\pi i/k}$. Denoting $x=J_{3}$, the one-form with values in the chiral ring is 

\begin{equation*}
dW(x)= \alpha\sum_{k=0}^{2} \frac{\zeta_{3}^{-k} dx}{x-x_{k}}
\end{equation*}

where $x_{0}=J_{3}(0)=0,x_{1}=J_{3}(1)=\zeta_{3}$ and $x_{2}=J_{3}(1/2)=1+\zeta_{3}$, with $x_{k+1}=T^{*}x_{k}$. The cusp at $\tau=\infty$ is instead sent to $x=\infty$ on the sphere. It is straightforward to check that $dW(x)$ satisfies $T^{*}dW(x)= \zeta_{3}^{-1}dW(x)$. \\ Another spherical version of a platonic solid appears for $N=4$. In this curve the cusps are the $6$ vertices of a regular octrahedron in $\mathbb{P}^{1}$ with symmetry group  $PSL(2,\mathbb{Z})/\Gamma(4)\sim \mathbb{S}_{4}$. Also in this case the co-levels are just $\pm 1$. The critical points are given by 

\begin{equation*}
C_{\Gamma(4)}=\left\lbrace 0,1/3,1/2,2/3,1,\infty\right\rbrace .
\end{equation*}

In this curve we have two UV cusps, i.e. $\infty$ and $1/2$, which are both fixed points of $T$. The remaining ones represent instead free IR critical points. The Hauptmodul of level $4$ is 

\begin{equation*}
J_{4}(\tau)=\zeta_{8}^{3}\sqrt{8}q_{\tau}^{1/4} \prod_{n=1}^{\infty} \frac{(1-q_{\tau}^{4n})^{2}(1-q_{\tau}^{n/2})}{(1-q_{\tau}^{n/4})^{2}(1-q_{\tau}^{2n})},
\end{equation*}

which transforms under $T$ as 

\begin{equation*}
T^{*}J_{4}= \frac{\zeta_{4}J_{4}}{1-J_{4}}.
\end{equation*}

The fixed points of this map are $0$ and $1-\zeta_{4}$, which correspond respectively to $\tau= \infty, \frac{1}{2}$. The $4$ IR cusps are all in the same orbit of $T$ and fall in the points $J_{4}(0)=\infty, J_{4}(1)=-\zeta_{4}, J_{4}(2/3)=1/(1+\zeta_{4}), J_{4}(1/3)=1$. These must be simple poles for $dW(x)$, with $x=J_{4}$, which reads

\begin{equation}
dW(x)= \alpha \left( \frac{1}{x-1} -\frac{1}{x+\zeta_{4}}+ \frac{\zeta_{4}}{x-1/(1+\zeta_{4})}\right)dx,
\end{equation}

and satisfies $T^{*}dW(x)=\zeta_{4}^{-1}dW(x)$. \\ $X(5)$ is the last case of genus $0$. This modular curve has $12$ cusps which identify the vertices of a regular icosahedron. The quotient $PSL(2,\mathbb{Z})/\Gamma(5)$ acts on the Riemann sphere as $\mathbb{A}_{5}$, the symmetry group of this platonic solid. Among the $12$ cusps

\begin{equation*}
C_{\Gamma(5)}=\left\lbrace 0,2/9,1/4,2/7,1/3,2/5,1/2,5/8,2/3,3/4,1,\infty \right\rbrace 
\end{equation*}

the UV fixed points are $2/5$ and $\infty$, while $\left\lbrace 0,2/9,1/4,3/4,1\right\rbrace $ and $\left\lbrace 2/7,1/3,1/2,5/8,2/3   \right\rbrace $ represent the $5$ decoupled vacua in two inequivalent IR limits. This curve has two inequivalent co-levels, i.e. $l=1,2$ and the Galois group acts on the set of cusps by exchanging the two UV limits and the two groups of IR theories. As in the previous cases, we can use as projective coordinate on $\mathbb{P}^{1}(\mathbb{C})$ the Hauptmodul 

\begin{equation*}
J_{5}(\tau)= \zeta_{5} q_{\tau}^{-1/5} \prod_{n=1}^{\infty} \frac{(1-q_{\tau}^{5n-2})(1-q_{\tau}^{5n-3}) }{(1-q_{\tau}^{5n-4})(1-q_{\tau}^{5n-1}) }
\end{equation*}

which transforms under $T$ as 

\begin{equation*}
T^{*}J_{5}= \zeta_{5}^{-1}J_{5}.
\end{equation*}

The UV cusps $2/5,\infty$ are sent by $J_{5}$ respectively to $x=0,\infty$, which are the fixed points of $T$. 
We know from \ref{dw} that a transformation $\gamma \in \Gamma_{0}(N)$ with $b=0$ mod $N$ acts on the cyclotomic units through the Galois group and permutes the residue of the poles in $dW(x)$. It follows that if two IR cusps are in the same orbit of such $\gamma$, their residue must be related by the correspondent Galois transformation. It is the case for instance of $0$ and $5/8$ if we set $b=5$ and $c=8$. Imposing also the correct transformation property under $T$ for $l=1$, one can repeat the procedure and fix all the coefficients up to an overall constant. Setting $x_{1;k}=\zeta_{5}^{k}J_{5}(0)$ and  $x_{2;k}=\zeta_{5}^{k}J_{5}(5/8)$, with $J_{5}(0)=1+\zeta_{5}+\zeta_{5}^{2}$ and $J_{5}(5/8)=(\zeta_{5}^{2}-\zeta_{5}^{4})/(1+\zeta_{5}-\zeta_{5}^{3}-\zeta_{5}^{4})$, we find

\begin{equation*}
dW(x)= \alpha \sum_{k=0}^{4} \left( \frac{1}{x-x_{1;k}} +  \frac{1}{x-x_{2;k}}\right)\zeta_{5}^{k} dx.
\end{equation*}

where the residue of $J_{5}(0)$ and $J_{5}(5/8)$ are both normalized to $1$.\\ By summarizing, the generators of loops around the IR cusps on the sphere correspond to the subset of generators of $\Gamma(N)$ which act non trivially on the critical value. Once the normalization is fixed, the constant generated by modular transformations can be computed with the residue formula. Moreover, the free IR cusps are all in the same orbit of $\Gamma_{0}(N)$, which acts by multiplication on the coefficients of $dW(x)$. This implies in particular that all the poles have the same order, which must be $1$ from the non trivial monodromy of $W(p_{0};x)$, and their residue are related by Galois transformations.\\ A similar procedure could be carried on in principle also for modular curves of higher genus, but it is more complicated. It is instead much more convenient to find a general expression of the critical value as function on the upper half plane and study its modular properties. Let us take the multi-valued function in \ref{symb}. Since the constants generated by the modular transformations are indipendent from the point, we can set $z=0$. The expression of $W^{(N,l)}(0;\tau)$ reads

\begin{equation}
W^{(N,l)}(0;\tau)= \sum_{k=0}^{N-1} e^{\frac{2\pi i l k}{N}} \log \left[ \Theta \begin{bmatrix} \frac{1}{2}-\frac{l}{N} \\ \frac{1}{2}-\frac{k}{N}\end{bmatrix}\left( 0;\tau \right) e^{- 2 \pi i \left( \frac{l}{N}-\frac{1}{2}\right)\left( \frac{k}{N}-\frac{1}{2}\right) }\right] .
\end{equation}

where the phase $e^{- 2 \pi i \left( \frac{l}{N}-\frac{1}{2}\right)\left( \frac{k}{N}-\frac{1}{2}\right)}$ is a convenient normalization constant. This function remains ill defined as long as we do not specify the determination of the logarithm. This is equivalent to choose, for a fixed $\tau$, a representative of $z=0$ and determine its critical value.  We first set the notations

\begin{equation*}
\begin{split}
 &  q_{\tau}= e^{2\pi i \tau}, \hspace{2cm} q_{z}= e^{2\pi i z}, \\ \\ & \hspace{1.3 cm} z= u_{1} \tau+ u_{2}, 
\end{split}
\end{equation*}

with $ u_{1},u_{2} \in \mathbb{Z}/ N $.  Then, we recall the definition of Siegel functions:

\begin{equation}\label{siegel}
g_{u_{1},u_{2}}(\tau)=- q_{\tau}^{B_{2}(u_{1})/2} e^{2\pi i u_{2}(u_{1}-1)/2}(1-q_{z}) \prod _{n=1}^{\infty} (1-q_{\tau}^{n}q_{z})(1-q_{\tau}^{n}/q_{z}),
\end{equation}

where $B_{2}(x)= x^{2}-x+\frac{1}{6}$ is the second Bernoulli polynomial. Because of their modular properties, these objects are a sort of `building blocks' for the modular functions of level $N$. In particular, all the Hauptmoduls defined previously can be expressed in terms of $ g_{u_{1},u_{2}}(\tau)$ and the Dedekind eta function \cite{rif7,rif8,rif13}.
The theta functions  $  \Theta \begin{bmatrix} \frac{1}{2}-u_{1} \\ \frac{1}{2}-u_{2}\end{bmatrix}\left( 0;\tau \right)$ have the $q$-product representation 

\begin{equation}
\begin{split}
\Theta \begin{bmatrix} \frac{1}{2}-u_{1} \\ \frac{1}{2}-u_{2}\end{bmatrix}\left( 0;\tau \right)= & - q_{\tau}^{B_{2}(u_{1})/2}q_{\tau}^{1/24}  e^{2\pi i (u_{1}-1/2)(u_{2}-1/2)} (1-q_{z}) \\ & \mathrm{x} \prod _{n=1}^{\infty}(1-q_{\tau}^{n}) (1-q_{\tau}^{n}q_{z})(1-q_{\tau}^{n}/q_{z})
\end{split}
\end{equation}

and can be written in terms of $g_{u_{1},u_{2}}(\tau) $  and $\eta (\tau) $ as 

\begin{equation}
 \Theta \begin{bmatrix} \frac{1}{2}-u_{1} \\ \frac{1}{2}-u_{2}\end{bmatrix}(0;\tau)= i g_{u_{1},u_{2}}(\tau) \eta (\tau) e^{ 2 \pi i u_{1}\left( u_{2}-1\right) /2}.
\end{equation}

Since the Siegel and Dedekind functions have neither zeroes nor poles, there is a single-valued branch of $\log \Theta \begin{bmatrix} \frac{1}{2}-u_{1} \\ \frac{1}{2}-u_{2}\end{bmatrix}\left( 0;\tau \right)$ on the upper half plane. Therefore, the critical value can be consistently defined as holomorphic function of $\tau \in \mathbb{H}$. Provided the above relations, we can rewrite $W^{(N,l)}(0;\tau)$ as  

\begin{equation}\label{mass}
W^{(N,l)}(0;\tau)= \sum_{k=0}^{N-1} e^{\frac{2\pi i l k}{N}} \log E_{\frac{l}{N},\frac{k}{N}}(\tau),
\end{equation}

where $E_{\frac{l}{N},\frac{k}{N}}(\tau)$ is the Siegel function of characters $ u_{1}=\frac{l}{N}, u_{2}=\frac{k}{N}$ without the root of unity $e^{2\pi i u_{2}(u_{1}-1)/2} $. Under an integer shift of the characters, these functions satisfy \cite{rif7}

\begin{equation}\label{shift}
E_{u_{1}+1,u_{2}}(\tau)=-e^{-2\pi i u_{2}} E_{u_{1},u_{2}}(\tau), \hspace{2cm} E_{u_{1},u_{2}+1}(\tau)=E_{u_{1},u_{2}}(\tau).
\end{equation}

Moreover, being the Siegel functions up to a multiplicative constant, they have good modular properties. For $\gamma= \begin{pmatrix}   a & b \\ c & d \end{pmatrix} \in SL(2,\mathbb{Z})$, they transform with a phase:

\begin{equation}\label{prop}
\begin{split}
& E_{u_{1},u_{2}}(\tau+b)=  e^{\pi i b B_{2}(u_{1})} E_{u_{1},u_{2}+b u_{1}}(\tau) , \ \ \ \ \mathrm{for } \ c = 0 , \\ \\ 
& E_{u_{1},u_{2}}(\gamma (\tau))= \varepsilon (a,b,c,d) e^{\pi i \delta} E_{u_{1}^{\prime},u_{2}^{\prime}}(\tau) , \ \ \ \ \mathrm{for } \ c\neq 0,
\end{split}
\end{equation}

where

\begin{equation}
\begin{split}
\varepsilon(a,b,c,d)= &
   \begin{cases}
  e^{i\pi (bd(1-c^{2})+c(a+d-3))/6}, \ \ \ \ \mathrm{if} \ c \ \mathrm{is \ odd},
   \\  -ie^{i\pi (ac(1-d^{2})+d(b-c+3))/6}, \ \ \ \ \mathrm{if} \ d \ \mathrm{is \ odd},
   \end{cases} \\ \\ 
    \delta= & u_{1}^{2}ab + 2 u_{1}u_{2}bc + u_{2}^{2}cd - u_{1}b-u_{2}(d-1),
    \end{split}
\end{equation}

and 

\begin{equation}\label{character}    
u_{1}^{\prime}= a u_{1}+c u_{2}, \hspace{1cm}  u_{2}^{\prime}= b u_{1}+d u_{2}.
\end{equation}

In order to compute the constants generated by the modular transformations, we need to evaluate the difference 

\begin{equation}\label{diff}
\chi_{u_{1},u_{2}}(\gamma)=\log E_{u_{1},u_{2}}(\gamma (\tau))-\log E_{u_{1}^{\prime},u_{2}^{\prime}}(\tau),
\end{equation}

for $\gamma \in SL(2,\mathbb{Z})$. Here we assume the characters of the Siegel functions to be normalized such that $0 < u_{1},u_{2},u_{1}^{\prime},u_{2}^{\prime} < 1$. This can always be achieved by the property \ref{shift}. The computation for the case of $\Gamma(N)$ has already been done in \cite{rif8}. In Appendix we follow closely that derivation, adapting it to the general case. For $c=0$ the transformations belong to the coset group $\Gamma_{1}(N)/\Gamma(N) \simeq \mathbb{Z}_{N}$ and we obtain 

\begin{equation}
\chi_{u_{1},u_{2}}(\gamma)= 2\pi i \frac{1}{2}B_{2}(u_{1}).
\end{equation}

It is clear that in this case we cannot appreciate a modular shift of the critical value. Indeed, these transformations simply translate the vacua:

\begin{equation}\label{trivial}
W^{(N,l)}(0;\tau+b)=e^{-\frac{2\pi i b l^{2}}{N}} W^{(N,l)}(0;\tau).
\end{equation}

On the other hand, for $c \neq 0$, we get the formula  

\begin{equation}\label{formula} 
\begin{split}
 \chi_{u_{1},u_{2}}(\gamma)= & \ 2\pi i \frac{1}{2}  \left( B_{2}(u_{1}) \frac{a}{c} + B_{2}(u_{1}^{\prime}) \frac{d}{c}  -\frac{2}{c} B_{1}(u_{1}^{\prime}) B_{1}( \langle d u_{1}^{\prime}- u_{2}^{\prime} c \rangle ) \right)  \\ \\ & -\frac{2 \pi i}{c}  \sum_{\overset{x \in \mathbb{Z}/c\mathbb{Z},}{ \ x \neq 0} }  [ x, u_{1}^{\prime}, u_{2}^{\prime} ]_{d,c},
 \end{split}
\end{equation}

where $B_{1}(x)=x-1/2$ is the first Bernoulli polynomial, $\langle x \rangle $ represents the fractional part of $x$ and the symbol $ [ x, u_{1}^{\prime}, u_{2}^{\prime} ]_{d,c}$ denotes

 \begin{equation}
 [ x, u_{1}^{\prime}, u_{2}^{\prime} ]_{d,c}=  \frac{e^{2\pi i x \bigl( \frac{ \langle d u_{1}^{\prime}-cu_{2}^{\prime}\rangle-du_{1}^{\prime} }{c}+u_{2}^{\prime} \bigr)}  }{(1-e^{-2\pi i x d/c})(1-e^{2\pi i x/c})}.
 \end{equation}

This result turns out to be indipendent from the branch of the logarithm and in particular from $\tau$. This is consistent with the fact that the modular shift of the critical value is indipendent from the vacuum that we choose. From the general formula we can reduce to the cases of the congruence subgroups. Let us consider $\gamma \in \Gamma_{0}(N)$. Using the fact that $c=0$ mod $N$ and $ad=1$ mod $N$, we have to plug in the above expression:

\begin{equation}
\begin{split}
u_{1}^{\prime}= \langle  a u_{1} \rangle ,&  \hspace{2cm} u_{2}^{\prime}=  \langle  d u_{2}+b u_{1}  \rangle, \\ \\
& \langle d u_{1}^{\prime}-u_{2}^{\prime}c  \rangle= u_{1}.
\end{split}
\end{equation}

If $\gamma \in \Gamma_{1}(N)$, these becomes 

\begin{equation}
\begin{split}
u_{1}^{\prime}= u_{1},&  \hspace{2cm} u_{2}^{\prime}=  \langle  u_{2}+ b u_{1}  \rangle, \\ \\
& \langle d u_{1}^{\prime}-u_{2}^{\prime}c  \rangle= u_{1}.
\end{split}
\end{equation}

The case of $\gamma \in \Gamma(N)$ follows from this by requiring further $b=0$ mod $N$.\\  Setting $u_{1}= \frac{l}{N}$, $u_{2}= \frac{k}{N}$ and summing over $k$ with the residue $e^{\frac{2\pi i l k}{N}}$, we find the modular transformations of the physical mass. In sequence

\begin{equation}\label{transf}  
\begin{split}
& \gamma \in \Gamma_{0}(N): \\ 
& W^{(N,l)}(0;\gamma(\tau))=  e^{-\frac{2\pi i al^{2}b}{N}} W^{(N,al)}(0;\tau) + \Delta W^{(N,l)}_{\Gamma_{0}(N)}(\gamma), \\ \\ 
  & \gamma \in \Gamma_{1}(N): \\ 
& W^{(N,l)}(0;\gamma(\tau))=  e^{-\frac{2\pi i l^{2}b}{N}} W^{(N,l)}(0;\tau) + \Delta W^{(N,l)}_{\Gamma_{1}(N)}(\gamma), \\ \\ 
& \gamma \in \Gamma(N): \\ 
& W^{(N,l)}(0;\gamma(\tau))=   W^{(N,l)}(0;\tau) + \Delta W^{(N,l)}_{\Gamma(N)}(\gamma), 
\end{split}
\end{equation}

with 

\begin{equation}\label{delta}
\begin{split}
& \Delta W^{(N,l)}_{\Gamma_{0}(N)}(\gamma)= -\frac{2 \pi i N}{c} \ e^{-\frac{2\pi i al^{2}b}{N}}\sum_{\overset{x \in \mathbb{Z}/c\mathbb{Z},}{ \ x= -al \ \mathrm{mod} \ N}} \frac{e^{2\pi i x \bigl( \frac{l/N-d 
\langle al/N \rangle }{c} \bigr)} }{(1-e^{-2\pi i x d/c})(1-e^{2\pi i x/c})}, \\ \\ 
& \Delta W^{(N,l)}_{\Gamma_{1}(N)}(\gamma)= -\frac{2 \pi i N}{c} \ e^{-\frac{2\pi i l^{2}b}{N}}\sum_{\overset{x \in \mathbb{Z}/c\mathbb{Z},}{ \ x= -l \ \mathrm{mod} \ N}} \frac{e^{2\pi i x \bigl( \frac{l}{N}\frac{1-d}{c} \bigr)  }}{(1-e^{-2\pi i x d/c})(1-e^{2\pi i x/c})},
\end{split}
\end{equation}

\begin{equation*}
\Delta W^{(N,l)}_{\Gamma(N)}(\gamma)= -\frac{2 \pi i N}{c} \sum_{\overset{x \in \mathbb{Z}/c\mathbb{Z},}{ \ x= -l \ \mathrm{mod} \ N}} \frac{e^{2\pi i x \bigl( \frac{l}{N}\frac{1-d}{c} \bigr)  }}{(1-e^{-2\pi i x d/c})(1-e^{2\pi i x/c})},
\end{equation*}

where the constraints on $x$ follow from the summation over $k$. These formulas are coherent with the transformations of $\partial_{z}W^{(N,l)}(z;\tau)$ that we found in the previous paragraph. \\ We can check in the simple case of $N=2$ that the formulas above give the results obained with the geometrical approach. The physical mass has the expression 

\begin{equation}
W(0;\tau)= \log \frac{\Theta_{4}(0;\tau)}{\Theta_{3}(0;\tau)} = \frac{1}{4}\log(1- \lambda(\tau)),
\end{equation}

where $1-\lambda(\tau)= \left( \Theta_{4}(0;\tau)/ \Theta_{3}(0;\tau) \right) ^{4}$ is the Hauptmodul of level $2$ that we have defined previously. Although $1-\lambda(\tau)$ is invariant under transformation of $\Gamma(2)$, the logarithm is not. We can use the expression for $\Delta W_{\Gamma(2)}$ to derive the modular transformations of $\log(1-\lambda(\tau))$. $\Gamma(2)$ is freely generated by the matrices 

\begin{equation}
T_{1}= \begin{bmatrix}
1 & 2 \\ 0 & 1
\end{bmatrix}, \hspace{2cm}   T_{2}= \begin{bmatrix}
1 & 0 \\ -2 & 1
\end{bmatrix} .
\end{equation}

Using respectively the \ref{trivial} and the \ref{transf}, \ref{delta} with $N=2, l=1$, one finds 

\begin{equation}
\begin{split}
& \log(1-\lambda(\tau+2))= \log(1-\lambda(\tau)), \\ \\ & \log\left( 1-\lambda\left( \frac{\tau}{1-2\tau}\right)\right)  = \log(1-\lambda(\tau))+2\pi i.
\end{split}
\end{equation}

It is clear that $T_{2}$ is the generator of anticlockwise loops around the IR cusp in $\tau=0$. Indeed, the constant is the same we obtain with the residue formula.

\subsection{Modular Transformations of the Ground State Metric}

The modular shift of the superpotential seems to contraddict the statement that the model is invariant under transformations of $\Gamma(N)$. But, if we assume the perspective of the universal cover, there is no contraddiction at all. Indeed, the physical mass parametrizes not only models, but also vacua. The modular transformation simply changes the initial choice of the vacuum $p_{0}$ with another one of the same fiber in the universal cover. Therefore, the new coordinate on the spectral curve describes the same model, but a different vacuum. \\ The $tt^{*}$ equations for these class of theories are manifestly covariant under transformations of the congruence subgroups. In particular, the covariance under $\Gamma_{1}(N)$ implies that the equation naturally descends on the space of models. However, as a consequence of the modular shift, the ground state metric is not left invariant. Indeed, matrices of $\Gamma(N)$ and $\Gamma_{1}(N)$ change the basis of lattice generators and consequently the representation of the homology group. Besides this effect, a transformation of $\Gamma_{0}(N)$ changes also the torsion point of the $\mathbb{Z}_{N}$ symmetry, resulting in a permutation of the metric components. Let us consider this more general case. From the transformation of $\partial_{z} W^{(N,l)}(z;\tau)$ under $\Gamma_{0}(N)$ one can read how the generators of the symmetry group are modified. We note that the new function is still periodic of $2\pi \tau$. Therefore, we can consider again $B$ as a generator of $H_{1}(\mathcal{S};\mathbb{Z}) $ in the new model.
The operators $L_{k}$ are left invariant by the transformation as well, since the corresponding homology cycles have the same definition in the model of co-level $al$. Differently, $\sigma$ and $A$ change in relation to the transformation of the torsion point of the vacua. For a $\gamma \in \Gamma_{0}(N)$, we can write

\begin{equation*}
\begin{split}
& \gamma^{*}B=B, \\ 
& \gamma^{*}L_{k}= L_{k} \\
& \frac{2\pi}{N} \longrightarrow \frac{2\pi}{N} (c\tau + d )  \Longrightarrow      \begin{cases} \gamma^{*}\sigma= \sigma^{d} B^{\frac{c}{N}}= \tilde{\sigma} B^{\frac{c}{N}} \\  \gamma^{*}A = A^{d} B^{c}= \tilde{A} B^{c},
\end{cases}
\end{split}
\end{equation*}

where we denote with $\tilde{\sigma}=\sigma^{d}$ the operator associated to the torsion point of co-level $al$, and with $\tilde{A}=A^{d}$ the homology cycle satisfying $\tilde{A}=\tilde{\sigma}^{N}$. The fact that the generator of loops are not involved in the modular transformations is consistence with the truncation of the $tt^{*}$ equation that we discussed in section $3$.\\ We have also to include the shift $z\rightarrow z -\frac{2\pi}{N}bl$, which implies the vacuum transformation

\begin{equation*}
p_{0} \longrightarrow \sigma^{-bl}(p_{0})  \Longrightarrow   \ket{p_{0}} \mapsto \sigma^{-bl}\ket{p_{0}} .
\end{equation*}

Let us study what these transformations mean at the level of vacuum states. We consider again trivial representations of $L_{k}$. The action of $\gamma$ on the point basis is 

\begin{equation*}
\gamma \ket{k;\alpha, \beta}=\  e^{\frac{-i \alpha k}{N}} \sum_{m,n \in \mathbb{Z}} e^{-i( m\alpha + n\beta )} (A^{d} B^{c})^{m} B^{n}  (\sigma^{d} B^{\frac{c}{N}})^{ k} \sigma^{-bl}\ket{p_{0}}
\end{equation*}

\begin{equation*}
\begin{split}
& = \ e^{-i\frac{kd}{N}\left( \frac{\alpha-\beta c }{d}\right) } \sum_{m,n\in \mathbb{Z}} e^{-i \left(  m \left( \frac{\alpha-\beta c}{d}\right)  + n\beta \right)} A^{m} B^{  n} \sigma^{kd-bl}\ket{p_{0}} \\  \\ & = \ e^{-i\frac{bl}{N} \left( \frac{\alpha-\beta c}{d} \right) } \ket{kd-bl; (\alpha-\beta c)/d, \beta}.
\end{split}
\end{equation*}

We note that, in the case of $ b=0$ mod $N$ and $d=1$ mod $N$, the fiber index $k$ is left invariant. This follows from the fact that a transformation of $\Gamma(N)$ preserves the torsion point up to a shift of a lattice vector, which moves the base point $p_{0}$ without changing the fiber. The $\sigma$-eigenstates transform consequently as 

\begin{equation*}
\gamma \ket{j;\alpha, \beta}=\  e^{-i\frac{bl}{N} \left( \frac{\alpha-\beta c}{d} \right) } \sum_{k=0}^{N-1}  e^{- \frac{2 \pi i l k j}{N}}  \ket{kd-bl;(\alpha-\beta c)/d , \beta} 
\end{equation*}

\begin{equation*}
\begin{split}
= & \ e^{-i\frac{bl}{N} \left(  2\pi a l j + \frac{\alpha-\beta c}{d}\right)} \sum_{k=0}^{N-1}  e^{- \frac{2 \pi i al k j}{N}}  \ket{k;(\alpha-\beta c)/d , \beta} \\ \\ 
= & \ e^{-i\frac{bl}{N} \left(  2\pi a l j + \frac{\alpha-\beta c}{d}\right)} \ket{aj;(\alpha-\beta c)/d, \beta}.
\end{split}
\end{equation*}

The action of the symmetry group generators $\tilde{\sigma},\tilde{A},B$ on the transformed states is given by 

\begin{equation*}
\begin{split}
& \tilde{\sigma}\ket{aj;(\alpha-\beta c)/d, \beta}= e^{\frac{2\pi i l j}{N}} e^{\frac{i}{N}(\alpha-\beta c)}\ket{aj;(\alpha-\beta c)/d, \beta}, \\ \\ 
& \tilde{A}\ket{aj;(\alpha-\beta c)/d, \beta}= e^{i(\alpha-\beta c)}\ket{aj;(\alpha-\beta c)/d, \beta}, \\ \\ 
& B\ket{aj;(\alpha-\beta c)/d, \beta}= e^{i\beta}\ket{aj;(\alpha-\beta c)/d, \beta}.
\end{split}
\end{equation*}

From the eigenvalues of the new basis, we learn that the ground state metric transforms in the following way:

\begin{equation}\label{metrictransformation}
\gamma^{*}\varphi_{j}(w;\alpha,\beta)= \varphi_{j}(\gamma^{*}w;\alpha,\beta)= \varphi_{aj}(w;\alpha-\beta c,\beta).
\end{equation}

As anticipated, we see that a transformation of $\Gamma(N)$ changes the representation of the homology, resulting in the character shift $\alpha\rightarrow \alpha-\beta c$. In the case of $\Gamma_{0}(N)$, since the operators $\tilde{\sigma}$ and $\sigma$ are inequivalent and have a different set of eigenstates, we appreciate also the exchange of the metric components along the diagonal. This effect takes place specifically for $N>2$, where we have a non trivial co-level structure. We note further that transformations of the coset $\Gamma_{1}(N)/\Gamma(N)$, i.e. with $c=0$, leave the metric completely invariant.

\section{Physics of the Cusps}

\subsection{Classification of the Cusps}

Now that we have discussed the modular properties of the solution, we want to describe its behaviour around the boundary regions of the domain. These are represented by the cusps of the modular curve $\mathbb{H}/\Gamma(N)$, i.e. the equivalence classes of $\Gamma(N)$ in $\mathbb{Q} \cup \lbrace\infty \rbrace$. First of all, we have to understand which type of model each cusp corresponds to. Let us begin with the cusp at $\tau= \infty$. It is convenient to come back to the initial lattices of poles and vacua with the shift $z\rightarrow z + \frac{2\pi l \tau}{N}$. Moreover, we rewrite the derivative of the superpotential in terms of $\Theta_{1}(z;\tau)$ as 

\begin{equation}\label{new}
\partial_{z} W^{(N,l)}(z;\tau)= \  \frac{1}{2} \sum_{k=0}^{N-1} e^{\frac{2\pi i lk}{N}} \frac{\Theta_{1}^{\prime}\left( \frac{1}{2}\left( z-\frac{2\pi k}{N}\right) ;\tau\right) }{\Theta_{1}\left(\frac{1}{2}\left( z-\frac{2\pi k}{N}\right) ;\tau\right)}. 
\end{equation}

Using the relation \cite{rif6}

\begin{equation*}
\frac{\Theta_{1}^{\prime}(z,\tau)}{\Theta_{1}(z ;\tau)}= \cot z 
+4 \sum_{n=1}^{\infty} \frac{q_{\tau}^{n}}{1-q_{\tau}^{n}}\sin 2nz,
\end{equation*}

with $q_{\tau}=e^{2\pi i \tau} $, the expression above becomes 

\begin{equation*}
\begin{split}
\partial_{z} W^{(N,l)}(z;\tau)= & \frac{1}{2} \sum_{k=0}^{N-1} e^{\frac{2\pi i lk}{N}} \biggl[ \cot\left(\frac{1}{2}\left( z-\frac{2\pi k}{N}\right) \right) +  \\ \\ & 4 \sum_{n=1}^{\infty} \frac{q_{\tau}^{n}}{1-q_{\tau}^{n}}\sin \left( 2n\left( \frac{1}{2}\left( z-\frac{2\pi k}{N}\right) \right) \right)  \biggr]. 
\end{split}
\end{equation*}

The theta function in these expressions is normalized with quasi-periods $\pi,\pi\tau$. Let us denote with $S$ the infinite sum in $n$. Manipulating the expression, we get 

\begin{equation*}
\begin{split}
S= & \ 2 \sum_{k=0}^{N-1} e^{\frac{2\pi i lk}{N}} \sum_{n=1}^{\infty} \frac{q_{\tau}^{n}}{1-q_{\tau}^{n}}\sin \left( n \left( z-\frac{2\pi k}{N}\right)\right)   \\ \\ = & -i \sum_{n=1}^{\infty} \frac{q_{\tau}^{n}}{1-q_{\tau}^{n}}  \left( e^{inz}\sum_{k=0}^{N-1} e^{\frac{2\pi i k}{N}(l-n)} - e^{-inz}\sum_{k=0}^{N-1} e^{\frac{2\pi i lk}{N}(l+n)} \right) .
 \end{split}
\end{equation*} 

The two sums over $k$ are not $0$ if and only if $n$ satisfies respectively $n=l$ mod $N$ and $n=-l$ mod $N$. Therefore, the derivative of the superpotential becomes 

\begin{equation*}
\begin{split}
\partial_{z} W^{(N,l)}(z;\tau) = & \ \frac{1}{2} \sum_{k=0}^{N-1} e^{\frac{2\pi i lk}{N}} \cot \left( \frac{1}{2}\left( z-\frac{2\pi k}{N}\right) \right)  \\ \\ & -iN \left(  \sum_{\overset{n=1}{n=l \ \mathrm{mod} \ N}}^{\infty} \frac{q_{\tau}^{n}}{1-q_{\tau}^{n}}  e^{inz}-  \sum_{\overset{n=1}{n=-l \ \mathrm{mod} \ N}}^{\infty}\frac{q_{\tau}^{n}}{1-q_{\tau}^{n}}  e^{-inz} \right). 
\end{split}
\end{equation*}

Taking the limit $\tau\rightarrow \infty$, the series in $q_{\tau}$ are truncated at the leading order. Moreover, since the vacua $-\frac{2\pi l \tau}{N} + \frac{2\pi k}{N} + \Lambda_{\tau}$ escape to infinity for large $\tau$, we have also to take $z \rightarrow  \infty$. Thus, we obtain 

\begin{equation*}
\partial_{z}W^{(N,l)}(z;\tau) \xrightarrow{ \tau\rightarrow \infty} -iN \left( q_{\tau}^{l } \ e^{ilz} - q_{\tau}^{(N-l)} \ e^{-i(N-l)z}  \right). 
\end{equation*}

Integrating this expression, we find 

\begin{equation}\label{limit}   
W^{(N,l)}(z;\tau) \xrightarrow{ \tau\rightarrow \infty} -N\left( q_{\tau}^{l }\  \frac{e^{ilz}}{l} + q_{\tau}^{(N-l)} \ \frac{e^{-i(N-l)z}}{N-l}  \right) ,
\end{equation}

wich we recognize as the superpotential of a $\hat{A}_{N-1}$ model of co-level $l$. In particular, the case of $N=2$ corresponds to the Sinh-Gordon model 

\begin{equation}\label{sinhgordon}
W(z;\tau) \sim  q_{\tau} \cos z .
\end{equation}

Now let us consider the rationals. We associate to a cusp $\frac{a}{c}$ with $\mathrm{gcd}(a,c)=1$ a modular transformation $\gamma_{\frac{a}{c}}=\begin{pmatrix} a & b \\ c & d \end{pmatrix}$ which sends $\infty$ to $\frac{a}{c}$. In this definition $b,d$ are integers such that $\gamma_{\frac{a}{c}} \in SL(2,\mathbb{Z})$ and clearly the case of $c=0$ corresponds to take again $\tau=\infty$. One can study the behaviour of the model around $ \tau=\frac{a}{c}$ by acting on $ \partial_{z} W^{(N,l)}(z;\tau)$ with $\gamma_{\frac{a}{c}}$ and then taking the limit $\tau \rightarrow \infty $. Using the modular properties \ref{weierstrass} of the zeta function, we have 

\begin{equation*}
\begin{split}
& \partial_{z} W^{(N,l)}\left( \frac{z}{c\tau + d};\frac{a\tau + b}{c\tau + d}\right) = \\ \\ &
(c\tau+d) \sum_{k=0}^{N-1} e^{\frac{2\pi i l k}{N}} \left[ \zeta \left( z-  \frac{2\pi}{N} k (c\tau + d) ;\tau \right) +2\frac{d \eta_{1}+c \eta_{2}}{N} k \right],
\end{split}
\end{equation*}

where the torsion point of the poles is now $\frac{2\pi}{N}(c\tau + d)$. Let us introduce the integers $Q= \mathrm{gcd} (c,N)$, with $1 \leq Q \leq \mathrm{min} \left\lbrace c,N \right\rbrace$, $j=\frac{N}{Q}$ and $r=\frac{c}{Q}$. Clearly we have $\mathrm{gcd} (r,j)=1$. By these definitions, we can split the sum over $k$ by writing $k=m+jp$, with $m=0,...,j-1$ and $p=0,...,Q-1$. Let us consider for the moment the cusps with divisor $Q>1$. One obtains

\begin{equation*}
\begin{split}
 & \partial_{z} W^{(N,l)}\left( \frac{z}{c\tau + d};\frac{a\tau + b}{c\tau + d}\right) = \\ \\ 
& (c\tau +d)\sum_{m=0}^{j-1} e^{\frac{2\pi i l m}{N}} \sum_{p=0}^{Q-1} e^{\frac{2\pi i l p}{Q}} \biggl[ \zeta \left( z-  \left( \frac{2\pi r}{j} \tau +\frac{2\pi d}{N} \right)m -\frac{2\pi}{Q} d p;\tau \right)  \\ \\ & + 2 \left( \frac{r\eta_{2} }{j} +  \frac{d\eta_{1} }{N}\right)m + \frac{2 d \eta_{1}}{Q}p \biggr].
\end{split}
\end{equation*}

As $\tau$ becomes very large, for $m\neq 0$ the torsion point approaches $\frac{2\pi r}{j}\tau$.\\  Moreover, by the formulas \cite{rif18}

\begin{equation*}
2\eta_{1}= \frac{G_{2}(\tau)}{2\pi}, \hspace{2 cm} 2\eta_{2}= \frac{\tau G_{2}(\tau)-2\pi i }{2\pi},
\end{equation*}

where $G_{2}(\tau)$ is the Eisenstein series 

\begin{equation*}
G_{2}(\tau)= \sum_{c ,d \in \mathbb{Z}\setminus \left\lbrace 0 \right\rbrace } \frac{1}{(c\tau +d)^{2}},
\end{equation*}

and the asymptotic behaviour 

\begin{equation}
G_{2}(\tau)\xrightarrow{ \tau\rightarrow \infty} 2\zeta(2),
\end{equation}

where $\zeta(z)$ is the Riemann zeta function, one has

\begin{equation}
\frac{\eta_{2}}{\eta_{1}} \xrightarrow{ \tau\rightarrow \infty}  \tau.
\end{equation}

Thus, we get the limit 

\begin{equation*}
\begin{split}
 & \partial_{z} W^{(N,l)}\left( \frac{z}{c\tau + d};\frac{a\tau + b}{c\tau + d}\right) \xrightarrow{ \tau\rightarrow \infty}  
 (c\tau +d) \sum_{p=0}^{Q-1} e^{\frac{2\pi i l p}{Q}} \biggl[ \zeta \left( z -\frac{2\pi}{Q} d p;\tau \right)+ \frac{2d\eta_{1}}{Q} p \biggr]_{\tau \rightarrow \infty} +    \\ \\ &  (c\tau +d)\sum_{m=1}^{j-1} e^{\frac{2\pi i l m}{N}} \sum_{p=0}^{Q-1} e^{\frac{2\pi i l p}{Q}} \biggl[ \zeta \left( z-   \frac{2\pi r}{j} \tau m ;\tau \right) + \frac{2 r\eta_{2}}{j}m \biggr]_{\tau \rightarrow \infty} = \\ \\ & (c\tau +d) \sum_{p=0}^{Q-1} e^{\frac{2\pi i l p}{Q}} \biggl[ \zeta \left( z -\frac{2\pi}{Q} d p;\tau \right)+ \frac{2d\eta_{1}}{Q} p \biggr]_{\tau \rightarrow \infty} = \\ \\ & (c\tau +d) \sum_{p=0}^{Q-1} e^{\frac{2\pi i al p}{Q}} \biggl[ \zeta \left( z -\frac{2\pi}{Q} p;\tau \right)+ \frac{2\eta_{1}}{Q}p \biggr]_{\tau \rightarrow \infty},
\end{split}
\end{equation*}

where in the last line we have used the fact that $ad=1$ mod $Q$. So, we learn from this expression and the limit  \ref{limit} that the cusp $\frac{a}{c}$ with divisor $Q=\mathrm{gcd} (c,N)$ correspond to an $\hat{A}_{Q-1}$ model of co-level $al$:

\begin{equation}\label{ahat}
W^{(N,l)}(z;\tau) \xrightarrow{ \tau\rightarrow \frac{a}{c}} -Q\left( q_{\tau}^{al }\  \frac{e^{ialz}}{al} + q_{\tau}^{(Q-al)} \ \frac{e^{-i(Q-al)z}}{Q-al}  \right),
\end{equation}

which is a theory with $Q$ vacua up to periodicity $z \sim z+2\pi$.\\
In the case of $Q=1$, the poles $\frac{2\pi k}{N}(c\tau +d) + \Lambda_{\tau}$ are all pushed to infinity when $\tau$ becomes large except for $k=0$. Thus, using the \ref{new} we can write symbolically

\begin{equation*}
W \sim `` \log \Theta_{1}(z/2;\tau\rightarrow \infty)  ",
\end{equation*}

and using the asymptotics 

\begin{equation*}
\Theta_{1}(z;\tau) \xrightarrow{ \tau\rightarrow \infty} 2 q_{\tau}^{\frac{1}{8}}\sin z,
\end{equation*}

one finds that these cusps are decribed by the multi-valued superpotential

\begin{equation}\label{free}
W(z)=`` \log \sin \left( z/2 \right) ".
\end{equation}

This model represents the free version of our class of theories, with a $1$ dimensional lattice of poles and one of vacua.\\ We know that the cusps of $X_{1}(N)$ are ramification points of the cover $X(N)\rightarrow X_{1}(N)$. Denoting with $\Gamma_{\frac{a}{c}}$ the stability group of the cusp $a/c$ in $\Gamma_{1}(N)$, one can write the equality \cite{rif2}

\begin{equation}\label{cond}
 \gamma_{\frac{a}{c}}^{-1}\Gamma_{\frac{a}{c}}\gamma_{\frac{a}{c}}= \left\langle \pm \begin{pmatrix} 1 & h \\ 0 & 1 \end{pmatrix}     \right\rangle,
\end{equation} 

which is satisfied with one of the two signs. This relation means that the generator of $\Gamma_{\frac{a}{c}}$ is conjugated to $\pm \begin{pmatrix} 1 & h \\ 0 & 1 \end{pmatrix}$ in $ \gamma_{\frac{a}{c}}^{-1} \Gamma_{1}(N) \gamma_{\frac{a}{c}} $. \\ The number $h$ is called width of the cusp and represents the minimal integer such that $a/c +h \sim a/c$ in $\Gamma(N)$. The absolute value can be equal or less than $N$ and count the number of degenerate vacua of the model labelled by the cusp $\left[ \frac{a}{c} \right]$ of $\Gamma_{1}(N)$. The cusps which satisfy the relation with the plus sign are called regular. From the theorems \ref{width1}, \ref{width2}, the stability condition can be written as:

\begin{equation*}  
\begin{bmatrix} a+ch \\ c \end{bmatrix}= \begin{bmatrix} a \\ c \end{bmatrix} \ \mathrm{mod} \ N.
\end{equation*}

It is clear that the minimal integer $h$ satisfying this condition is $h=j=\frac{N}{Q}$. Thus, for the cusps with divisor $Q$, the $N$ vacua split in $j$ decoupled theories which appear on $X(N)$ as $Q$-degenerate points. These are described by the $\hat{A}_{Q-1}$ models \ref{ahat} for $1<Q\leq N$, or by the free theories \ref{free} if $Q=1$. In particular, the cusps with $Q=N$, or equivalently $c=0$ mod $N$, are the UV limits, since the vacua tend to a unique point on the spectral curve. \\ The exceptions to this picture are represented by the so called irregular cusps, i.e. those which satisfy the \ref{cond} with the mignus sign. In this case the stabilizer of the cusp belongs to $-\Gamma_{1}(N)$ and we have

\begin{equation*}
\begin{bmatrix} a+ch \\ c \end{bmatrix}= -\begin{bmatrix} a \\ c \end{bmatrix} \ \mathrm{mod} \ N.
\end{equation*}

If we exclude the trivial case of $N=2$ where $1\sim-1$ and require $a$ and $c$ to be coprime, we find that the stability condition is satisfied only by cusp $1/2$ for the curve of level $4$. This is known to be the unique irregular cusp for $\Gamma_{1}(N)$. Despite we have $Q=N/Q=2$, the width of the cusp is $h=1$, and the corresponding theory is actually a $\hat{A}_{3}$ model with $4$ vacua. The superpotential is the Sinh-Gordon one in \ref{sinhgordon} as for the cusps with divisor $2$, but we have to impose the identification $z \sim z+ 4\pi$.\\
Now we want to determine the positions of the cusps on the W-plane. Using the expression \ref{mass} for the critical value, we have 

\begin{equation*}
\begin{split}
W^{(N,l)}(0;\tau\rightarrow \frac{a}{c})=& \ W^{(N,l)}(0;\gamma_{\frac{a}{c}}(\tau\rightarrow \infty))= 
\sum_{k=0}^{N-1} e^{\frac{2\pi i l k}{N}} \log E_{\frac{l}{N},\frac{k}{N}}(\gamma_{\frac{a}{c}}(\tau\rightarrow \infty)) \\ \\=  &   \sum_{k=0}^{N-1} e^{\frac{2\pi i l k}{N}} \log E_{\big<\frac{al+ck}{N}\big>,\big<\frac{dk+bl}{N}\big>}(\tau\rightarrow \infty) + \sum_{k=0}^{N-1} e^{\frac{2\pi i l k}{N}} \chi_{\frac{l}{N};\frac{k}{N}}(\gamma_{\frac{a}{c}}),
\end{split}
\end{equation*}

where $ \chi_{\frac{l}{N};\frac{k}{N}}(\gamma_{\frac{a}{c}})$ is given by the formula in \ref{formula}. \\ Let us consider the limit of the first piece. The leading order of $E_{u_{1},u_{2}}(\tau)$ for $\tau \rightarrow \infty$ is 

\begin{equation*}
\mathrm{ord}_{i \infty} E_{u_{1},u_{2}}(\tau)= \frac{1}{2} B_{2}\left( \langle u_{1} \rangle \right) .
\end{equation*}

Therefore, we find 

\begin{equation*}
\begin{split}
W^{(N,l)}(0;\tau\rightarrow \frac{a}{c})= & \sum_{k=0}^{N-1} e^{\frac{2\pi i l k}{N}} \log E_{\big<\frac{al+ck}{N}\big>,\big<\frac{dk+bl}{N}\big>}(\tau)
\xrightarrow{ \tau\rightarrow \infty}  \sum_{k=0}^{N-1} e^{\frac{2\pi i l k}{N}} \log q_{\tau}^ {\frac{1}{2} B_{2}\left( \big<\frac{al+ck}{N} \bigr> \right) }  \\ \\ =& \log q_{\tau}^{K^{(N,l)}_{\frac{a}{c}}},
 \end{split}
\end{equation*}

where 

\begin{equation*}
{K^{(N,l)}_{\frac{a}{c}}}= \frac{1}{2}\sum_{k=0}^{N-1} e^{\frac{2\pi i l k}{N}} B_{2}\left( \bigg<\frac{al+ck}{N}\bigg> \right). \hspace{2cm} 
\end{equation*}

Let us develop this expression. Using the Fourier expansion of the second Bernoulli periodic polynomial 

\begin{equation*}
B_{2}(\langle x \rangle)= -\frac{2!}{(2\pi i)^{2}} \sum_{\overset{m=-\infty}{m \neq 0}}^{\infty} \frac{e^{2\pi i m x}}{m^{2}},
\end{equation*}

we get 

\begin{equation*}
\begin{split}
{K^{(N,l)}_{\frac{a}{c}}}= & -\frac{1}{(2\pi i)^{2}} \sum_{k=0}^{N-1} e^{\frac{2\pi i l k}{N}} \sum_{\overset{m=-\infty}{m \neq 0}}^{\infty} \frac{e^{2\pi i m \left( \frac{al+ck}{N}\right) }}{m^{2}} \\ \\  = & -\frac{1}{(2\pi i)^{2}} \sum_{\overset{m=-\infty}{m \neq 0}}^{\infty} \frac{e^{2\pi i m \frac{al}{N}}}{m^{2}} \sum_{k=0}^{N-1} e^{\frac{2\pi i k}{N}(l+mc)}.
\end{split}
\end{equation*}

The sum $\sum_{k=0}^{N-1} e^{\frac{2\pi i k}{N}(l+mc)} $ is not vanishing if and only if $l+mc=0$ mod $N$, which admits solution only when $c$ is coprime with $N$. Thus, we have 

\begin{equation*}
{K^{(N,l)}_{\frac{a}{c}}}= \begin{cases} 
-\frac{N}{(2\pi i)^{2}}^{-\frac{2\pi i a l^{2}r}{N}}  \sum\limits_{\overset{m=-\infty}{m= -lr \ \mathrm{mod} \ N}}^{\infty} \frac{1}{m^{2}} \neq 0, \ \ \ \mathrm{if} \ \mathrm{gcd}(c,N)=1, \\ \\ 
0, \ \ \  \mathrm{otherwise}, \end{cases}
\end{equation*}

where $r= c^{-1}$ mod $N$. So, we learn that if $\mathrm{gcd}(c,N)=1$ the cusp order is not vanishing and therefore the critical value is divergent. Coherently with our analysis, these are the IR fixed points described by free theories. In this limits all the vacua decouple and the solitons connecting them become infinitely massive. On the contrary, the cusps with $1< Q \leq N $ have a finite coordinate on the W-plane:

\begin{equation*}
W^{(N,l)} \left( 0;\frac{a}{c}\right)=\sum_{k=0}^{N-1} e^{\frac{2\pi i l k}{N}} \chi_{\frac{l}{N};\frac{k}{N}}(\gamma_{\frac{a}{c}}).
\end{equation*}

In the case of $c=0$ mod $N$ this becomes 

\begin{equation*}
W^{(N,l)} \left( 0;\frac{a}{c}\right)^{\mathrm{UV}}= \Delta W^{(N,l)}_{\Gamma_{0}(N)}(\gamma_{\frac{a}{c}}).
\end{equation*}

In particular, the cusp $\tau= \infty$ is located at the origin and provides the boundary condition for the critical limit $\mu\rightarrow 0$. \\ It is clear from the last formula that the UV cusps are all in the same orbit of the Galois group of the modular curve. Counting the equivalence classes of $\Gamma_{1}(N)$, these models are labelled by the co-levels $\pm l$ and their number is equal to $\phi(N)/2$. From this point of view, choosing the co-level is equivalent to pick which UV cusp to put in the origin. Also the free IR cusps are all in the same orbit of the Galois group, since we can map $\tau=0$ to a generic rational $a/c$ such that $\mathrm{gcd}(c,N)=1$ with a matrix of $\Gamma_{0}(N)$. Instead, the other IR cusps with divisor $1< Q < N$ can split in different equivalence classes of $\Gamma_{0}(N)$. In general the Galois group maps a cusp $a/c$ with $\mathrm{gcd}(c,N)=Q$ and $\mathrm{gcd}(a,Q)=1$ to another cusp $\frac{a^{\prime}}{c^{\prime}}$ with $\mathrm{gcd}(c^{\prime},N)=Q$ and $\mathrm{gcd}(a^{\prime},Q)=1$. A computation in \cite{rif2} shows that for a given divisor $Q$ we have $\phi(\mathrm{gcd}(Q,N/Q))$ cusps of $\Gamma_{0}(N)$.

\subsection{Local Solutions}

\subsubsection{Boundary Conditions}

In this last section we provide the boundary conditions for the $tt^{*}$ equations and describe the solution around the cusps. Approaching a critical point, the solution has to match the asymptotic behavour required by the physics of the corresponding cusp. The deformations in the space of couplings near these points regard the overall parameter $\mu$ rescaling the superpotential. Since an overall phase can always be absorbed in the fermionic measure of the superspace, the solution must depends only on the module $\vert \mu \vert$. As we said, the RG fixed point is reached when $\mu\rightarrow 0$. It is known that at the critical point a generic Landau-Ginzburg theory gains the $U(1)_{V}$ R-symmetry, which is broken off-criticality by the superpotential. In this limit the $tt^{*}$ connection is shown to approach the matrix $Q$ which generates the R-symmetry \cite{rif4,rif10,rif11}. Denoting with $t=\vert \mu \vert $, one can define 

\begin{equation*}
Q_{i\bar{j}}(t)=\frac{1}{2}(gt\partial_{t}g^{-1})_{i,\bar{j}}+\frac{\hat{c}}{2}, \hspace{2cm}  \lim_{t\rightarrow 0} Q_{i\bar{j}}(t)=Q_{i\bar{j}},
\end{equation*}

where $\hat{c}$ is the central charge of the critical theory. In this sense, the Berry's connection associated to the RG flow can be seen as an off-criticality definition of the $U(1)$ generator. Near the UV fixed point the ground state metric can be diagonalized in a basis of vacua with definite charge:

\begin{equation*}
g_{i\bar{i}} \xrightarrow{ \mu \rightarrow 0} (\mu\bar{\mu})^{-(q_{i}-\hat{c}/2)}
\end{equation*}

where the charges $q_{i}$ are real numbers distributed among $q_{\mathrm{min}}=0$ and $q_{\mathrm{max}}=\hat{c}$. So, the solution to the equations near the critical point is given in terms of the charges of the $U(1)_{V}$ eigenstates.\\ In what follows we study the $U(1)$ spectrum of the UV and IR cusps.

\subsubsection{UV Cusps}

The $\hat{A}_{N-1}$ models are well studied in literature \cite{rif10}. They are Landau-Ginzburg theories with superpotential 

\begin{equation*}
W^{(N,l)}(z;t)= \mu \left( \frac{e^{-lz}}{l} +  \frac{e^{(N-l)z}}{N-l}\right) ,
\end{equation*}

and $\mathrm{gcd}(l,N)=1$. These are integrable models with a $\mathbb{Z}_{N}$ symmetry generated by $\sigma: z\rightarrow z + \frac{2\pi i}{N}$ and $N$ vacua, provided the periodic identification $z \sim z+2\pi i$. The symmetry acts transitively on the critical points, which are given by the condition $e^{Nz}=1$. In the UV limit these theories tend to $\sigma$-models over abelian orbifolds of $\mathbb{CP}^{1}$. These are known to be asymptotically free theories with central charge $\hat{c}=1$ \cite{rif10}. The $tt^{*}$ equations in canonical form are the Toda equations in \ref{toda} with vanishing $\beta$. Indeed, in this limit the unique non trivial homology operator is $A=\sigma^{N}$ and the $U(1)$ charges defining the solution can depend only on $\alpha$. Since $U(1)$ is broken by the superpotential to $\mathbb{Z}_{N}$, it is clear that the generators of the two symmetry groups have a common basis of eigenstates. Let us first consider the case of $\alpha=0$. A basis of $U(1)$ eigenstates in the chiral ring can be generated with the operators $e^{-z}, e^{z}$. Since the superpotential must have R-charge $1$, these have respectively charge $\frac{1}{l}$ and $\frac{1}{N-l}$. Given that $e^{-lz}= e^{(N-l)z}$ in the chiral ring from the vacua condition, we find that the set of eigenstates split in two `towers'

\begin{equation*}
\begin{split}
 & e^{-z} \hspace{1cm} e^{-2z} \hspace{1cm}\cdot \hspace{1cm} \cdot \hspace{1cm} \cdot \hspace{1cm} \cdot \hspace{1cm} \cdot \hspace{1cm} e^{-(l-1)z} \\ 
&  e^{z} \hspace{1.2cm} e^{2z} \hspace{1.25cm} \cdot \hspace{1cm} \cdot \hspace{1cm} \cdot \hspace{1cm} \cdot \hspace{1cm} \cdot \hspace{1cm} e^{(N-l-1)z}  
\end{split}
\end{equation*}

with $U(1)_{V}$ charges 

\begin{equation*}
\begin{split}
 & \hspace{0.45cm} \frac{1}{l} \hspace{1.8cm} \frac{2}{l}\hspace{1.5cm} \cdot \hspace{1cm} \cdot \hspace{1cm} \cdot \hspace{1cm} \cdot \hspace{1cm} \cdot \hspace{1.5cm} \frac{l-1}{l} \\ \\ 
&  \frac{1}{N-l} \hspace{1cm} \frac{2}{N-l}\hspace{1.13cm} \cdot \hspace{1cm} \cdot \hspace{1cm} \cdot \hspace{1cm} \cdot \hspace{1cm} \cdot \hspace{1cm} \frac{N-l-1}{N-l}.
\end{split}
\end{equation*}

Approximately, we can say that the theory splits in two, with a set of operators dominant on the other one according to how we take the limit. We complete the basis by adding the identity $I$ and $e^{-lz}$, which have respectively charge $0$ and $1$. Near the critical point these two operators correspond to a unique marginal degree of freedom which gets a logarithmic correction to the scaling \cite{rif10}. We point out that in this language the Galois group acts directly on the $U(1)$ charges with the map $l\rightarrow al$ and puts in relation the solutions of the different $\hat{A}_{N-1}$ models. To see that the set of charges is invariant under this map we have to use the chiral ring condition $e^{Nz}=1$. The operatorial equality $e^{-kz}=e^{(N-k)z}$ for a generic $k \in \mathbb{Z}$ implies the equivalence $\frac{k}{l} \sim \frac{N-k}{N-l}$ at the level of corresponding charges. In general, the integer $k$ and the co-level $l$ are periodic of $N$ in the chiral ring. So, one can recast all the charges above as $q_{k}=\frac{k}{l}$, $k=0,...,N-1$ and write the action of the Galois group as $q_{k}\rightarrow\frac{k}{al}$. The relation $\frac{k}{l}=\frac{ak}{al}\sim\frac{k^{\prime}}{al}$, with $k^{\prime}= ak$ mod $N$, shows that the set of charges is left invariant by this map.\\ We can include the dependence from the angle $\alpha$ by multiplying the basis above by $e^{\frac{\alpha}{2\pi}z}$. In this way the operators have the correct eigenvalues under $\mathbb{Z}_{N}$ when $\alpha \neq 0$. The $U(1)$ charges as functions of the angle are 

\begin{equation*}
\begin{split}
 & \frac{1}{l}\left( k-\frac{\alpha}{2\pi}\right) , \ \ \ k=1,...,l-1, \\ \\ 
  & \frac{1}{N-l}\left( k+\frac{\alpha}{2\pi}\right), \ \ \ k=1,...,N-l-1, \\ \\ 
 & \frac{\alpha}{2\pi(N-l)}, \hspace{1cm} 1-\frac{\alpha}{2\pi l} .
\end{split}
\end{equation*}

It is clear from \ref{metrictransformation} that for $\beta=0$ a transformation of $\Gamma_{0}(N)$ does not change the dependence on $\alpha$ of the metric components. This can be seen at the level of charges by the fact that the map $l\rightarrow al$ is compensated by the rescaling of the angle $\alpha\rightarrow\alpha/d$.\\ We note further that, since $\beta$ is vanishing, the UV cusps turn out to be fixed points of $\Gamma(N)$. This is consistent with the fact that $A$ is the unique generator of the homology in this regime. \\ The irregular cusp $1/2$ of the modular curve of level $4$ is described by the superpotential

\begin{equation*}
W(z)= \mu \left( e^{2z} + e^{-2z} \right) 
\end{equation*}

with the identification $z \sim z+2\pi i$. This theory has a $Z_{2}$ symmetry generated by $\sigma: z\rightarrow z+\frac{i\pi}{2}$, but $4$ vacua determined by the condition $e^{2z}=e^{-2z}$. This model belongs to $\hat{A}_{3}$ family and is asymptotically a $\sigma$-model on the $\mathbb{CP}^{1}/\mathbb{Z}_{2} $ orbifold. The $tt^{*}$ equations are the Toda ones with $N=4$ and a basis of $U(1)$ eigenstates is given by 

\begin{equation*}
 e^{\frac{\alpha}{2\pi}z} \hspace{1cm} e^{\left( 1+\frac{\alpha}{2\pi} \right) z} \hspace{1cm} e^{-\left( 1-\frac{\alpha}{2\pi}\right) z} \hspace{1cm} e^{-\left( 2-\frac{\alpha}{2\pi} \right) z}
\end{equation*}

with charges respectively

\begin{equation*}
\frac{\alpha}{4\pi} \hspace{1cm} \frac{1}{2}\left(  1+\frac{\alpha}{2\pi}\right) \hspace{1cm} 
\frac{1}{2}\left(  1-\frac{\alpha}{2\pi} \right) \hspace{1cm} \frac{1}{2}\left(  2-\frac{\alpha}{2\pi} \right) . 
\end{equation*}

We conclude by saying that the solution of the $tt^{*}$ equation is singular in the UV cusps:

\begin{equation}
\varphi_{i}(t;\alpha) \xrightarrow{t\rightarrow 0} -2\left( q_{i}(\alpha) -\frac{1}{2}\right) \log t.
\end{equation}

A solution in terms of regular trascendents can be given only on the upper half plane, which is a simply connected space.

\subsubsection{IR Cusps}

The discussion for the $\hat{A}_{Q-1}$ models for $1< Q < N$ is pretty much the same of the previous paragraph. So, we focus on the free massive theories corresponding to the case of $Q=1$. These IR cusps are Landau-Ginzburg models described by the derivative

\begin{equation}
\partial_{z}W(z;\tau)= \mu \cot\left(  \frac{z}{2}  \right). 
\end{equation}

This function is periodic of $2\pi$ and has simple poles and simple zeroes respectively in $ 2k\pi $ and $\pi + 2k\pi$, $\k \in \mathbb{Z}$. Moreover, it is odd with respect to the parity transformation $\iota: z\rightarrow -z$. Since the target space is not simply connected we need to pull-back the model on the abelian universal cover. A natural basis for the homology is given by the cycles $B,B^{\prime}$ in figure \ref{algebr}. From the residue formula and the parity properties of $\partial_{z}W(z)$ one gets the transformations of the superpotential

\begin{equation}
\begin{split}
& B^{*}W(p)= W(p) - 2\pi i \mu, \\ 
& B^{\prime *}W(p)= W(p) + 2\pi i \mu.
\end{split}
\end{equation}

Proceeding as in \ref{point} we can construct the unique theta-vacua of this theory: 

\begin{equation}
\ket{\phi, \psi}= \sum_{n,m \in \mathbb{Z}} e^{-i(m\phi + n\psi)} B^{m}B^{\prime n} \ket{0},
\end{equation}

where we denote with $\ket{0}$ some vacuum state of the covering model. Setting to $0$ the corresponding critical value, the whole set is simply

\begin{equation}
W_{n,m}= 2\pi i \mu (n-m).
\end{equation}

We want to derive the $tt^{*}$ equation in the parameter $\mu$. The chiral ring operator $C_{\mu}(\phi, \psi)$ acts on the theta-vacuum as differential operator in the angles 

\begin{equation}
C_{\mu}\ket{\phi, \psi}=  \sum_{n,m \in \mathbb{Z}} e^{-i(m\phi + n\psi)}2\pi i(n-m) B^{m}B^{\prime n} \ket{0}=
2\pi \left( \frac{\partial}{\partial \phi}-\frac{\partial}{\partial \psi}\right)  \ket{\phi, \psi}.
\end{equation}

We define the ground state metric 

\begin{equation}
g(t,\phi, \psi)= \braket{\overline{\phi, \psi}}{\phi, \psi} = e^{L(t ,\phi,\psi)},
\end{equation}

where $L(t,\phi,\psi)$ is a real function of the angles and the RG scale $t=\vert \mu \vert$. We can normalize the state so that the topological metric is $1$. Thus, the reality constraint implies

\begin{equation}\label{reality}
\begin{split}
& L(-\phi,-\psi)=-L(\phi,\psi). 
\end{split}
\end{equation}

Moreover, by the commutation relations 

\begin{equation}
\begin{split}
& \iota B=B^{\prime -1}\iota, \\ \\ 
& \iota B^{\prime}=B^{-1} \iota, 
\end{split}
\end{equation}

we have also

\begin{equation}\label{parity}
L(-\psi,-\phi)=L(\phi,\psi).
\end{equation}

The $tt^{*}$ equation for $g(t,\phi, \psi)$ reads

\begin{equation}
\left( \partial_{\mu}\partial_{\bar{\mu}}+ 4\pi^{2}\left( \frac{\partial}{\partial \phi}-\frac{\partial}{\partial \psi}\right) ^{2}\right)  L(t,\phi,\psi)=0.
\end{equation}

We recognize in this expression the equation of a $U(1)$ Bogomolnyi monopole on $\mathbb{R}^{2}\times S^{1}$. Abelian $tt^{*}$ geometries in $(\mathbb{R}^{2}\times S^{1})^{r}$ have been studied in \cite{rif12,rif19}. The solution can be expanded in Bessel-MacDonald functions as 

\begin{equation}
L(t,\phi,\psi)= \sum_{ m_{1},m_{2} \in \mathbb{Z} \setminus \left\lbrace 0 \right\rbrace } A(m_{1},m_{2}) K_{0}\left( 4\pi t \vert m_{1}+m_{2} \vert \right)  \exp \left(  i \left(  m_{1}\phi-m_{2}\psi \right)  \right),
\end{equation}

where the coefficients $A(m_{1},m_{2})$ can be determined by imposing appropriate boundary conditions.
One can easily see that the $tt^{*}$ reality constraint \ref{reality} implies

\begin{equation}
A(-m_{1},-m_{2})=- A(m_{1},m_{2}), \hspace{1cm} A(m_{1},m_{2}) \in i \mathbb{R},
\end{equation}

while the parity condition \ref{parity} requires 

\begin{equation}
A(m_{2},m_{1})=A(m_{1},m_{2}).
\end{equation}

Combining these two conditions one gets the further constraint

\begin{equation}
L(t, \psi, \phi)=-L(t,\phi,\psi).
\end{equation}

According to the discussion in section \ref{trunc}, in order to have the abelianity of the solution one should consider trivial representations of the loop generator. If we demand the loop angle to vanish, namely $\phi=\psi$,  we simply find the trivial solution 

\begin{equation}
g(t,\phi, \phi)=1.
\end{equation}

\section{Conclusions}

In this paper we have shown how the $tt^{*}$ geometry of the modular curves is rich of interesting phenomena and outstanding connections between geometry, number theory and physics. These Riemann surfaces parametrize a family of supersymmetric FQHE models in which the usual setting degenerates in a doubly periodic physics on the complex plane. In the subclass of theories of level $N$, the elliptic functions playing the role of superpotentials have $N$ vacua and $N$ poles in the fundamental cell, with the corresponding residues which add up to zero by definition. The cancellation of the total flux between the magnetic field and the quasi-holes guarantees the enhancement of symmetry that makes possible to face analitically these models. In particular, the presence of an abelian symmetry group with a transitive action on the vacua allows to diagonalize the ground state metric, as well as to find the necessary topological data to write the $tt^{*}$ equations. This requires to pull-back the model on the abelian universal cover of the target manifold, where we have seen that the physics is non-abelian. On this space the symmetry group is enlarged with the generators of loops around the poles, which are responsible for the non trivial commutation relations between the generators of the algebra. Hovewer, the abelianity that we have required in the classification can be recovered at the quantum level. In particular, the ansatz of a solution with vanishing loop angles is consistent with all the $tt^{*}$ equations, which can be recasted as Toda equations in the canonical coordinates. \\ Studying the modular properties of these models, we have underlined that the non trivial modular transformations of the superpotential are a natural consequence of the geometry of the modular curves. A critical value as coordinate on the spectral cover can be defined only on the upper half plane, since the F-term variations are rational functions in projective coordinates on the modular curves. This has been studied in the easiest cases of the platonic solids inscribed in the Riemann sphere, but for surfaces of higher genus it is more convenient to parametrize the critical value in terms of the fundamental units of the modular function field. The congruence subgroups have a not trivial effect also on the components of the ground state metric, since they change the representation of the abelian symmetry group.\\ The known results and theorems about the cusps counting and classification have been recoverd in a physical language when we have classified the critical limits of this family of theories. One of the main point is that the width of a cusp allows to determine the UV or IR nature of the corresponding RG fixed point.\\  Our investigation has also revealed the algebraic properties of the modular curves. As we pointed out, the most remarkable connection with number theory is that the Galois group of the real cyclotomic extensions acts on the regularity conditions of the $\hat{A}_{N-1}$ Toda equations. This follows from the fact that the $\hat{A}_{N-1}$ models play the role of UV critical limits and belong to the same orbit of the Galois group.

\section*{Acknowledgements}

I want to thank my PhD advisor Sergio Cecotti for his useful guide and constant supervision of this work.

\appendix

\section{Modular Transformations of $\log E_{u_{1},u_{2}}(\tau)$}

In section $4$ we have setted the notations

\begin{equation*}
\begin{split}
 &  q_{\tau}= e^{2\pi i \tau}, \hspace{2cm} q_{z}= e^{2\pi i z}, \\ \\ & \hspace{1.3 cm} z= u_{1} \tau+ u_{2}, 
\end{split}
\end{equation*}

with $ u_{1},u_{2} \in \mathbb{Z}/ N $, and defined the modular units 

\begin{equation}\label{expan}  
E_{u_{1},u_{2}}(\tau)= q_{\tau}^{B_{2}(u_{1})/2} (1-q_{z}) \prod _{n=1}^{\infty} (1-q_{\tau}^{n}q_{z})(1-q_{\tau}^{n}/q_{z}),
\end{equation}

which are the Siegel functions up to the root of unity $e^{2\pi i u_{2}(u_{1}-1)/2}$. These objects satisfy \cite{rif7}

\begin{equation}
E_{u_{1}+1,u_{2}}(\tau)=-e^{-2\pi i u_{2}} E_{u_{1},u_{2}}(\tau), \hspace{2cm} E_{u_{1},u_{2}+1}(\tau)=E_{u_{1},u_{2}}(\tau),
\end{equation}

and transform under $\gamma=  \begin{pmatrix}   a & b \\ c & d \end{pmatrix} \in SL(2,\mathbb{Z})$ as 

\begin{equation}\label{prop1}
\begin{split}
& E_{u_{1},u_{2}}(\tau+b)=  e^{\pi i b B_{2}(u_{1})} E_{u_{1},u_{2}+b u_{1}}(\tau) , \ \ \ \ \mathrm{for } \ c = 0 , \\ \\ 
& E_{u_{1},u_{2}}(\gamma (\tau))= \varepsilon (a,b,c,d) e^{\pi i \delta} E_{u_{1}^{\prime},u_{2}^{\prime}}(\tau) , \ \ \ \ \mathrm{for } \ c\neq 0,
\end{split}
\end{equation}

where 

\begin{equation}
\varepsilon(a,b,c,d)= 
   \begin{cases}
  e^{i\pi (bd(1-c^{2})+c(a+d-3))/6}, \ \ \ \ \mathrm{if} \ c \ \mathrm{is \ odd},
   \\  -ie^{i\pi (ac(1-d^{2})+d(b-c+3))/6}, \ \ \ \ \mathrm{if} \ d \ \mathrm{is \ odd},
   \end{cases}
   \end{equation}
   
 \begin{equation*}
  \delta=  u_{1}^{2}ab + 2 u_{1}u_{2}bc + u_{2}^{2}cd - u_{1}b-u_{2}(d-1),
\end{equation*}

and 

\begin{equation}  
u_{1}^{\prime}= a u_{1}+c u_{2}, \hspace{1cm}  u_{2}^{\prime}= b u_{1}+d u_{2}.
\end{equation}

With these definitions, we want to compute the difference 

\begin{equation}
\chi_{u_{1},u_{2}}(\gamma)=\log E_{u_{1},u_{2}}(\gamma (\tau))-\log E_{u_{1}^{\prime},u_{2}^{\prime}}(\tau),
\end{equation}

for $\gamma \in SL(2,\mathbb{Z})$ and generic characters $u_{1},u_{2} \in \mathbb{Z}/N$. From \ref{prop1} we know that there is a power of $E_{u_{1},u_{2}}(\gamma (\tau))/E_{u_{1}^{\prime},u_{2}^{\prime}}(\tau)$ which is equal to one. This number is $12 N$ for $\Gamma(N)$ and $12 N^{2}$ for $\Gamma_{1}(N),\Gamma_{0}(N)$ and the whole $SL(2,\mathbb{Z})$. Therefore, the difference $\chi_{u_{1},u_{2}}(\gamma)$  must be equal to $2\pi i$ times a rational number. Given that the upper half plane is simply connected, this number is independent of $\tau$. Moreover, since $\log E_{u_{1},u_{2}}(\tau) $ is single-valued on the upper half plane, it is also indipendent from the branch of the logarithm. A natural choice, suggested by the $q$-expansion of the Siegel functions, is the principal branch on $\mathbb{C}$ with the negative real axis deleted. From now on we will use this determination. Because $E_{u_{1},u_{2}}(\tau)$ changes by a phase under an integer shift of the characters, we can  assume without loss of generality the canonical normalization  $0 < u_{1},u_{2},u_{1}^{\prime},u_{2}^{\prime} < 1$. \\ Let us first consider the case with $c=0$. These transformations belong to the coset group $\Gamma_{1}(N)/\Gamma(N) \simeq \mathbb{Z}_{N}$ and are generated by $\gamma (\tau)= \tau+1$. Using the expansion of the Siegel function in \ref{expan} we easily obtain 

\begin{equation}
\chi_{u_{1},u_{2}}(\gamma)= 2\pi i \frac{1}{2}B_{2}(u_{1}).
\end{equation}

From now on we assume $c\neq 0$ and write $\gamma(\tau)= \frac{a\tau +b}{c\tau +d}= \frac{a}{c}-\frac{1}{c^{2}\tau + cd}$. \\ Using again the \ref{siegel} we have 

\begin{equation*}
 \log E_{u_{1},u_{2}}(\tau)= 2\pi i B_{2}(u_{1}) \tau + \log(1-q_{z}) + \sum_{n=1}^{\infty} \left( \log(1-q_{\tau}^{n}q_{z})+
 \log(1-q_{\tau}^{n}/q_{z})\right).
\end{equation*}

With $\tau$ in the upper half plane and the characters canonically normalized, the conditions of absolute convergence for the standard series of the principal logarithm are satisfied. Therefore, using series expansions like

\begin{equation*}
\log(1-q_{z})=-\sum_{m=1}^{\infty} \frac{q_{z}^{m}}{m}
\end{equation*}

for the logarithms in the expression, we obtain 

\begin{equation*}
 \log E_{u_{1},u_{2}}(\tau)= 2\pi i \frac{1}{2}B_{2}(u_{1})\tau - \mathcal{Q}(z;\tau),
\end{equation*}

where 

\begin{equation*}
\mathcal{Q}(z;\tau)= \sum_{m=1}^{\infty} \frac{1}{m} \frac{q_{z}^{m}+ (q_{\tau}/q_{z})^{m}}{1-q_{\tau}^{m}}.
\end{equation*}

Then, let us put 

\begin{equation*}
\tau= -\frac{d}{c}+i y, \ \mathrm{with} \ y>0, \hspace{2cm} \gamma(\tau)= \frac{a}{c} + \frac{i}{c^{2}y}.
\end{equation*}

Since it is indipendent of $\tau$, we can calculate $ \chi_{u_{1},u_{2}}(\gamma)$ in the limit $y\rightarrow 0$, i.e. $\tau \rightarrow -\frac{d}{c}$ and $\gamma(\tau) \rightarrow i \infty$, by applying the Abel limit formula. 
Setting 

\begin{equation*}
z_{\gamma}= u_{1} \gamma(\tau)+u_{2}, \hspace{2cm} z^{\prime}= u_{1}^{\prime}\tau + u_{2}^{\prime},
\end{equation*}

and keeping only the immaginary parts, since $ \chi_{u_{1},u_{2}}(\gamma)$ is pure immaginary, we have to evaluate the expression 

\begin{equation}
\begin{split}
 \chi_{u_{1},u_{2}}(\gamma)=&\  2\pi i \frac{1}{2} \left( B_{2}(u_{1}) \frac{a}{c} +   B_{2}(u_{1}^{\prime}) \frac{d}{c}   \right) 
 \\ \\ &- \lim_{\tau\rightarrow -\frac{d}{c}}\left(  \mathrm{Im} \mathcal{Q}(z_{\gamma};\gamma({\tau}))-  \mathrm{Im} \mathcal{Q}(z^{\prime};\tau)\right) .
 \end{split}
\end{equation}

Let us start with $\mathrm{Im} \mathcal{Q}(z_{\gamma};\gamma({\tau}))$. As $\gamma(\tau) \rightarrow i \infty$, $q_{z_{\gamma}}$ and $q_{\gamma(\tau)}/q_{z_{\gamma}}$ approach $0$, therefore

\begin{equation*}
\lim_{\tau\rightarrow -\frac{d}{c}} \  \mathrm{Im} \mathcal{Q}(z_{\gamma};\gamma({\tau}))=0.
\end{equation*}

Now it is the turn of  $\mathcal{Q}(z^{\prime};\tau)$. We can decompose it in two pieces :

\begin{equation*}
\begin{split}
\lim_{\tau\rightarrow -\frac{d}{c}}\ \mathrm{Im}  \mathcal{Q}(z^{\prime};\tau)= & \ \lim_{\tau\rightarrow -\frac{d}{c}} \ \mathrm{Im} \sum_{c \ \nmid \ m} \frac{1}{m} \mathcal{Q}_{m}(z^{\prime};\tau) \  +  \ \lim_{\tau\rightarrow -\frac{d}{c}} \ \mathrm{Im} \sum_{c \mid m} \frac{1}{m}\mathcal{Q}_{m}(z^{\prime};\tau) \\ = & \ L^{\prime} +  L^{\prime \prime} ,
\end{split}
\end{equation*}

where 

\begin{equation*}
\mathcal{Q}_{m}(z^{\prime};\tau)=\frac{q_{z^{\prime}}^{m}+ (q_{\tau}/q_{z^{\prime}})^{m}}{1-q_{\tau}^{m}}.
\end{equation*} 

The symbols $L^{\prime}$ and $L^{\prime \prime}$ denote the sum respectively for $c \nmid m$ and $c \mid m $. We introduce

\begin{equation*}
r= e^{-2\pi y}, \hspace{1.5cm} M=N\vert c \vert, \hspace{1.5cm} \zeta= e^{-2\pi i d/c}, \hspace{1.5cm} \lambda= e^{2\pi i\left(  -\frac{d}{c}u_{1}^{\prime} + u_{2}^{\prime}   \right)}. 
\end{equation*}

It is shown in \cite{rif8},\cite{rif9} that the partial sums of these series are uniformly bounded. Therefore, we are allowed to take the limit under the sign of summation. Let us consider first $L^{\prime \prime}$. Using the notation above and taking the immaginary part, we have 

\begin{equation*}
L^{\prime \prime}= \lim_{r\rightarrow 1} \sum_{ c \mid m} \frac{r^{u_{1}^{\prime}m}-r^{(1-u_{1}^{\prime})m}}{1-r^{m}} \frac{1}{2m}(\lambda^{m}-\lambda^{-m}). 
\end{equation*}

Taking the limit under the summation sign, one gets

\begin{equation*}
\begin{split}
L^{ \prime \prime }= & \sum_{ c \mid m}  (1-2u_{1}^{\prime}) \frac{\lambda^{m}-\lambda^{-m}}{2m}= \sum_{m=1}^{\infty}(1-2 u_{1}^{\prime}) \frac{1}{2\vert c \vert m}( \lambda^{\vert c \vert m}-\lambda^{-\vert c \vert m}) \\ \\  = &
\ (1-2u_{1}^{\prime})\frac{1}{2 \vert c \vert}\sum_{m=1}^{\infty}\frac{1}{m}\left(  e^{2\pi i (-d \varepsilon(c)u_{1}^{\prime}+ u_{2}^{\prime} \vert c \vert )m} - e^{-2\pi i (-d \varepsilon(c)u_{1}^{\prime}+ u_{2}^{\prime} \vert c \vert )m} \right),
\end{split}
\end{equation*}

where $ \varepsilon (c)= \vert c \vert / c $. If $t$ is real and not integer, it holds the Fourier expansion 

\begin{equation*}
\sum_{m=1}^{\infty} \frac{1}{m}(e^{2\pi i m t}- e^{-2\pi i m t})= -2\pi i B_{1}\left( \langle t \rangle \right) ,
\end{equation*}

where  $ B_{1}(x)=x-\frac{1}{2}$ is the first Bernoulli polynomial. Thus 

\begin{equation*}
L^{ \prime \prime }= -2\pi i (1-2u_{1}^{\prime})\frac{1}{2 \vert c \vert} B_{1}\left( \langle -d \varepsilon(c)u_{1}^{\prime}+ u_{2}^{\prime} \vert c \vert \rangle\right) = -\frac{2\pi i}{c} B_{1}\left( u_{1}^{\prime}\right)   B_{1}\left(  \langle d u_{1}^{\prime}- u_{2}^{\prime} c \rangle \right) . 
\end{equation*}

Now we turn to the last piece $L^{\prime}$. Taking the limit under the summation sign, we obtain 

\begin{equation*}
L^{\prime}=\lim_{\tau\rightarrow -\frac{d}{c}} \ \mathrm{Im} \sum_{c \ \nmid \  m} \frac{1}{m}\mathcal{Q}_{m}(z^{\prime};\tau) = \sum_{c \ \nmid \ m} \frac{1}{m} \varphi (m),
\end{equation*}

where 

\begin{equation*}
\varphi(m)= \mathcal{Q}_{m}(z^{\prime};\tau)\vert_{\tau= -d/c} = \frac{\lambda^{m}+ (\zeta/\lambda)^{m}}{1-\zeta^{m}}.
\end{equation*}

Since $ \varphi(-m)=-\varphi(m)= \overline{\varphi(m)}$, we note that $\varphi(m)$ is pure immaginary and an odd function of $m$ mod $M=N \vert c \vert$. Now, for each class $x \in \mathbb{Z}/M\mathbb{Z}$ and $2x \not \in M\mathbb{Z}$, we define  

\begin{equation}
f(x)= \sum_{m=1}^{\infty} \frac{a(m,x)}{m}
\end{equation}

where

\begin{equation}
a(m,x)= \begin{cases}
0 \hspace{0.8cm} \mathrm{if} \  m \neq \pm x \ \mathrm{mod} \ M
\\ 1  \hspace{0.8cm} \mathrm{if} \ m = x \ \mathrm{mod} \ M
\\ -1  \hspace{0.5cm} \mathrm{if} \ m =- x \ \mathrm{mod} \ M.
\end{cases}
\end{equation}

Then, $L^{\prime}$ can be rewritten as 

\begin{equation}
L^{\prime}= \frac{1}{2}\sum_{\overset{x \in \mathbb{Z}/M\mathbb{Z},}{ \overset{\ x \neq 0 \ \mathrm{mod} \ c\mathbb{Z},}{ \ 2x \not \in M\mathbb{Z}} }}\varphi(x) f(x).
\end{equation}

In \cite{rif8} is shown that 

\begin{equation}
f(x)= \frac{-i \pi}{M}\left[  \frac{1}{1-e^{2\pi i x/M}} - \frac{1}{1-e^{-2\pi i x/M}} \right].
\end{equation}

Let $\omega= e^{2\pi i /N \vert c \vert}$. Using this expression $L^{\prime}$ becomes 

\begin{equation*}
\begin{split}
L^{\prime}= & \frac{-\pi i}{2M}\sum_{c \  \nmid \ x}\frac{\lambda^{x}+(\zeta/\lambda)^{x}}{1-\zeta^{x}}\left[  \frac{1}{1-\omega^{x}}-\frac{1}{1-\omega^{-x}} \right] \\ \\  
= & \frac{-\pi i}{2M}\sum_{c \ \nmid \ x}   \biggl[ \frac{\lambda^{x}}{(1-\zeta^{x})(1-\omega^{x})} +  \frac{(\zeta / \lambda)^{x}}{(1-\zeta^{x})(1-\omega^{x})}  \\ \\ & -  \frac{\lambda^{x}}{(1-\zeta^{x})(1-\omega^{-x})} - \frac{(\zeta / \lambda)^{x}}{(1-\zeta^{x})(1-\omega^{-x})}  \biggr].                                          
\end{split}
\end{equation*}

Changing $x$ to $-x$ in the last two terms, we find 

\begin{equation*}
L^{\prime}= -\frac{\pi i}{M} \left[    \sum_{c \ \nmid \ x}  \frac{\lambda^{x}}{(1-\zeta^{x})(1-\omega^{x})} + \sum_{c \ \nmid \ x}  \frac{(\zeta / \lambda)^{x}}{(1-\zeta^{x})(1-\omega^{x})} \right] .
\end{equation*}

This expression can be further simplified. We decompose the sum by introducing the variable

\begin{equation*}
\begin{split}
& x= y + k \vert c \vert, \\ \\
0 < y < \vert c \vert , & \hspace{2cm} 0 \leqslant k \leqslant N-1.
\end{split}
\end{equation*}

Let us denote with $S$ the partial sum in the variable $k$ of the first term in $L^{\prime}$. One gets

\begin{equation*}
\begin{split}
S= & \frac{\lambda^{y}}{1-\zeta^{y}}\sum_{k=0}^{N-1} \frac{\lambda^{k \vert c \vert}}{1-\omega^{y+k \vert c \vert}} \\ \\
 = & -\frac{1}{M}\frac{\lambda^{y}}{1-\zeta^{y}} \sum_{r=0}^{N-1}r\omega^{ry} \sum_{k=0}^{N-1} (\lambda \omega^{r})^{k \vert c \vert}.
\end{split}
\end{equation*}

The sum on the right is $0$ unless $ (\lambda \omega^{r})^{\vert c \vert}=1$. Using the definitions of $\lambda$ and $\omega$ in terms of $u_{1}^{\prime}, u_{2}^{\prime}, d,c$, we see that $(\lambda \omega^{r})^{\vert c \vert}=1$ if and only if 

\begin{equation*}
r=  Nd u_{1}^{\prime}-Ncu_{2}^{\prime} \ \ \mathrm{mod} \ N.
\end{equation*}

Letting consequently $r=N  \langle d u_{1}^{\prime}-cu_{2}^{\prime}\rangle + sN$ with $ 0 \leq s \leq \vert c \vert -1$, we have 

\begin{equation*}
\begin{split}
S & =  -\frac{1}{\vert c \vert} \frac{\lambda^{y}}{1-\zeta^{y}} \sum_{\overset{0 \leq r \leq N-1,}{ r= N u_{1}^{\prime}d-Ncu_{2}^{\prime} \ \mathrm{mod} \ N}}  r\omega^{ry} \\ \\ & =   -\frac{1}{\vert c \vert} \frac{\lambda^{y}}{1-\zeta^{y}} \sum_{s=0}^{\vert c \vert -1}(N  \langle d u_{1}^{\prime}-cu_{2}^{\prime}\rangle + sN)e^{2\pi i \frac{y}{Nc} (N  \langle d u_{1}^{\prime}-cu_{2}^{\prime}\rangle + sN)} \\ \\ & =  -\frac{1}{\vert c \vert} \frac{\lambda^{y}}{1-\zeta^{y}}\  e^{2\pi i \frac{y}{c} \langle d u_{1}^{\prime}-cu_{2}^{\prime}\rangle}  \sum_{s=0}^{\vert c \vert -1} s N e^{2\pi i ys/c}\\ \\ & =
N  \frac{\lambda^{y}}{1-\zeta^{y}} e^{2\pi i \frac{y}{c} \langle d u_{1}^{\prime}-cu_{2}^{\prime}\rangle}  \frac{1}{1-e^{2\pi i y/c}} \\ \\ & = N \frac{e^{2\pi i y \bigl( \frac{ \langle d u_{1}^{\prime}-cu_{2}^{\prime}\rangle-du_{1}^{\prime} }{c}+u_{2}^{\prime} \bigr)}  }{(1-e^{-2\pi i y d/c})(1-e^{2\pi i y/c})}. 
\end{split}
\end{equation*} 

In order to write the final result in a more compact way, we introduce the symbol 
 
 \begin{equation}
 [ x, u_{1}^{\prime}, u_{2}^{\prime} ]_{d,c}=  \frac{e^{2\pi i x \bigl( \frac{ \langle d u_{1}^{\prime}-cu_{2}^{\prime}\rangle-du_{1}^{\prime} }{c}+u_{2}^{\prime} \bigr)}  }{(1-e^{-2\pi i x d/c})(1-e^{2\pi i x/c})}.
 \end{equation}

Noting that the second sum in $L^{\prime} $ can be obtained from the first one with the substitution $ u_{1}^{\prime}\rightarrow 1- u_{1}^{\prime}, u_{2}^{\prime} \rightarrow -u_{2}^{\prime}$, we  get 

\begin{equation*}
L^{\prime}= -\frac{\pi i}{c} \left[  \sum_{\overset{x \in \mathbb{Z}/c\mathbb{Z},}{ x \neq 0 } } [ x, u_{1}^{\prime}, u_{2}^{\prime} ]_{d,c}    + \sum_{\overset{ x \in \mathbb{Z}/c\mathbb{Z},}{ x \neq 0} }  [ x, 1-u_{1}^{\prime}, -u_{2}^{\prime} ]_{d,c} \right] .
\end{equation*}

From the property $[ -x,1- u_{1}^{\prime},- u_{2}^{\prime} ]_{d,c}=[ x, u_{1}^{\prime}, u_{2}^{\prime} ]_{d,c} $, one obtains further 

\begin{equation}
L^{\prime}= -\frac{2 \pi i}{c} \sum_{\overset{x \in \mathbb{Z}/c\mathbb{Z},}{  x \neq 0 } } [ x, u_{1}^{\prime}, u_{2}^{\prime} ]_{d,c}  .
\end{equation}

Putting all the pieces together, we finally have 

\begin{equation}
\begin{split}
 \chi_{u_{1},u_{2}}(\gamma)= & \ 2\pi i \frac{1}{2} \left( B_{2}(u_{1}) \frac{a}{c} +   B_{2}(u_{1}^{\prime}) \frac{d}{c}  -\frac{2}{c} B_{1}(u_{1}^{\prime}) B_{1}( \langle d u_{1}^{\prime}- u_{2}^{\prime} c \rangle ) \right)  \\ \\ & -\frac{2 \pi i}{c}  \sum_{\overset{x \in \mathbb{Z}/c\mathbb{Z},}{ \ x \neq 0 }}  [ x, u_{1}^{\prime}, u_{2}^{\prime} ]_{d,c}.
 \end{split}
\end{equation}

\end{document}